\documentclass{article}

\usepackage{arxiv}

\usepackage[utf8]{inputenc} % allow utf-8 input
\usepackage[T1]{fontenc}    % use 8-bit T1 fonts
\usepackage{hyperref}       % hyperlinks
\usepackage{url}            % simple URL typesetting
\usepackage{booktabs}       % professional-quality tables
\usepackage{amsfonts}       % blackboard math symbols
\usepackage{nicefrac}       % compact symbols for 1/2, etc.
\usepackage{microtype}      % microtypography
\usepackage{lipsum}
\usepackage{graphicx}

\usepackage[utf8]{inputenc}
\usepackage[T1]{fontenc}
\usepackage{mathrsfs}
\usepackage{subcaption}
\usepackage[normalem]{ulem}
\usepackage{comment}
\usepackage{hyperref}
\usepackage{float}
\usepackage{amsmath}
\usepackage{amssymb}

\graphicspath{ {./images/} }

\title{Importance of Overlapping Network Nodes in Influence Spreading}

\author{
Kosti Koistinen \\
  Aalto University School of Science \\ Computer Science Department \\ P.O.Box 11000, 00076 \\ AALTO, Finland \\
  \texttt{kosti.koistinen@aalto.fi}
   \And
Vesa Kuikka \\
  Aalto University School of Science \\ Computer Science Department \\ P.O.Box 11000, 00076 \\ AALTO, Finland \\
  \texttt{vesa.kuikka@aalto.fi}
  \And
  Kimmo K. Kaski \\
  Aalto University School of Science \\ Computer Science Department \\ P.O.Box 11000, 00076 \\ AALTO, Finland \\
  \texttt{kimmo.kaski@aalto.fi}
}

\begin{document}
\maketitle
\begin{abstract}
In complex networks, “circles” are attribute-defined subgraphs whose nodes share common characteristics (e.g., group membership or categories). Nodes that belong to multiple such circles form overlapping regions, but their role in influence spreading processes remains somewhat underexplored. We analyse several networks with circle structures using a probabilistic influence spreading model for processes of simple and complex contagion in them. We quantify the importance of these overlapping nodes using three metrics, i.e., In-Centrality, Out-Centrality, and Betweenness Centrality, which represent the susceptibility, spreading power, and mediating role of nodes, respectively. We find that, at each stage of the spreading process, the overlapping nodes systematically exhibit greater influence than the non-overlapping nodes, even when accounting for structural heterogeneity, i.e., node connectivity. Furthermore, we observe that the criteria used to define circles shape the overlapping effects. When we restrict our analysis to only the largest circles, we find that circles reflect not only node-level attributes but also of topological importance. These findings help clarify the distinction between local attribute-driven circles and global community structures, thus highlighting the strategic importance of overlapping nodes in spreading dynamics. This provides a foundation for future research on overlapping nodes in both circles and communities.
\end{abstract}
\flushbottom
\maketitle
% * <kosti.koistinen@aalto.fi> 2015-02-09T12:07:31.197Z:
\thispagestyle{empty}
\section*{Introduction}
Network Science provides a powerful and flexible framework to investigate the properties and phenomena of natural and human-made systems, with applications that span from analysing and modelling social networks, epidemic spreading, cybersecurity,  and beyond \cite{newman2010networks, barabasi2016network}. By representing the entities of these systems as networks of nodes and relationships between them as edges, the network approach allows us to explore structural patterns, information flow, and dynamic processes in complex systems. One of the fundamental tasks in Network Science is community detection, which aims to identify densely connected subgraphs in complex networks \cite{fortunato2010community}. Communities are groups of nodes with stronger connections to each other than to other nodes of the network \cite{fortunato2016community}. In the context of social networks, these structures can be further refined into more granular units, often referred to as circles, which represent groups of nodes sharing common attributes or affiliations \cite{mcauley2012learning}. In this work, we define a \textit{circle} as an attribute-induced subgraph, i.e., a subset of nodes sharing a common attribute or group membership. These circles are not required to form cliques, triads or strictly dense subgraphs, although they may exhibit higher internal connectivity than the network average. Importantly, this definition differs from purely topological communities, as circles are constructed from node attributes rather than inferred solely from network structure. For example, a node representing an individual may belong to family, hobby, school, and work circles. Moreover, in real-world social networks, it is common for nodes to participate in multiple circles, which leads to overlapping structures \cite{mcauley2012learning}. From a broader perspective, these circles provide a useful framework to understand social interaction and epidemic dynamics, as both information and infections propagate through the contact patterns they represent \cite{pastor2015epidemic}.

\begin{figure}[ht]
  \centering
  \begin{subfigure}[t]{0.48\linewidth}
    \centering
    \includegraphics[width=\linewidth]{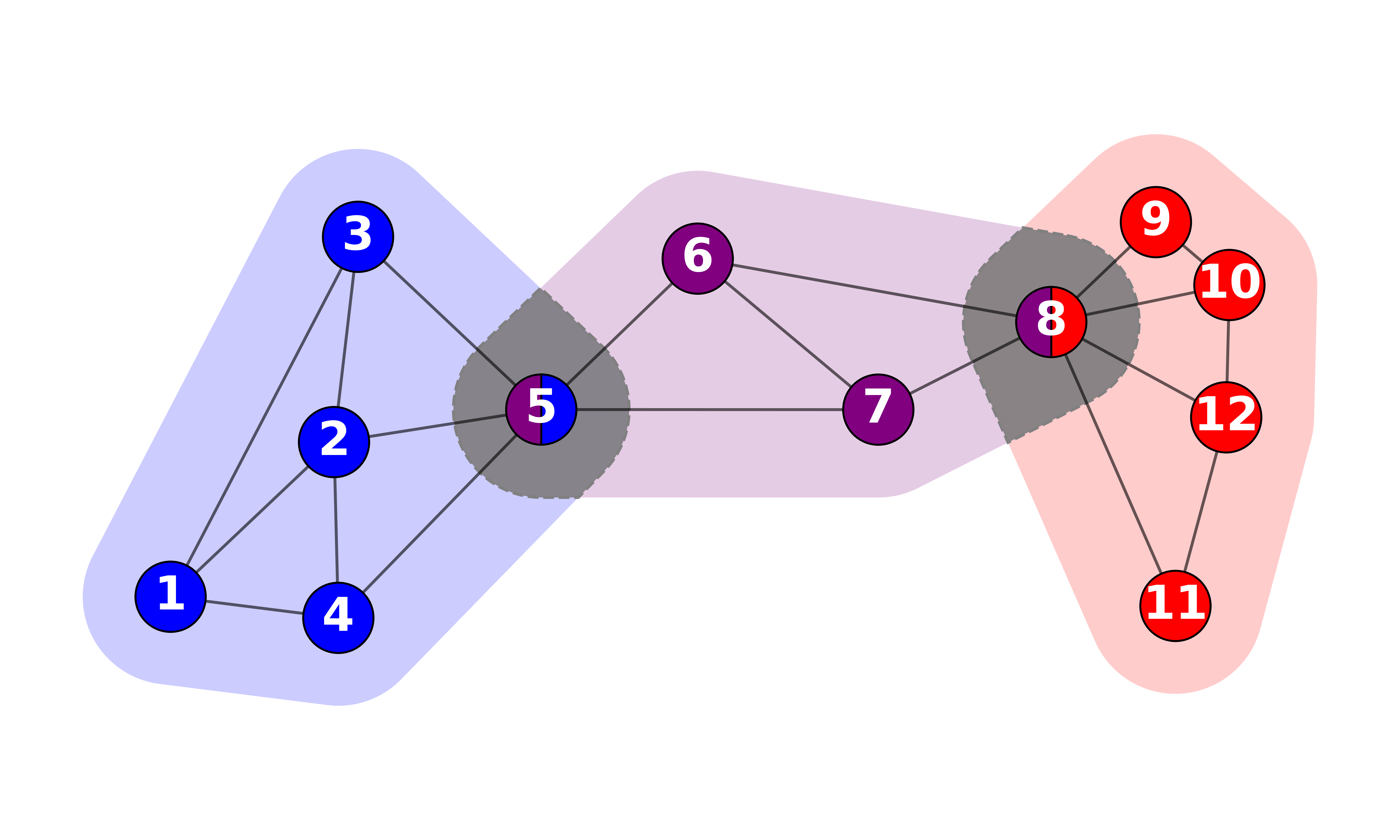}
    \caption{Overlapping Circles}
    \label{fig:illc}
  \end{subfigure}
  \hfill
  \begin{subfigure}[t]{0.48\linewidth}
    \centering
    \includegraphics[width=\linewidth]{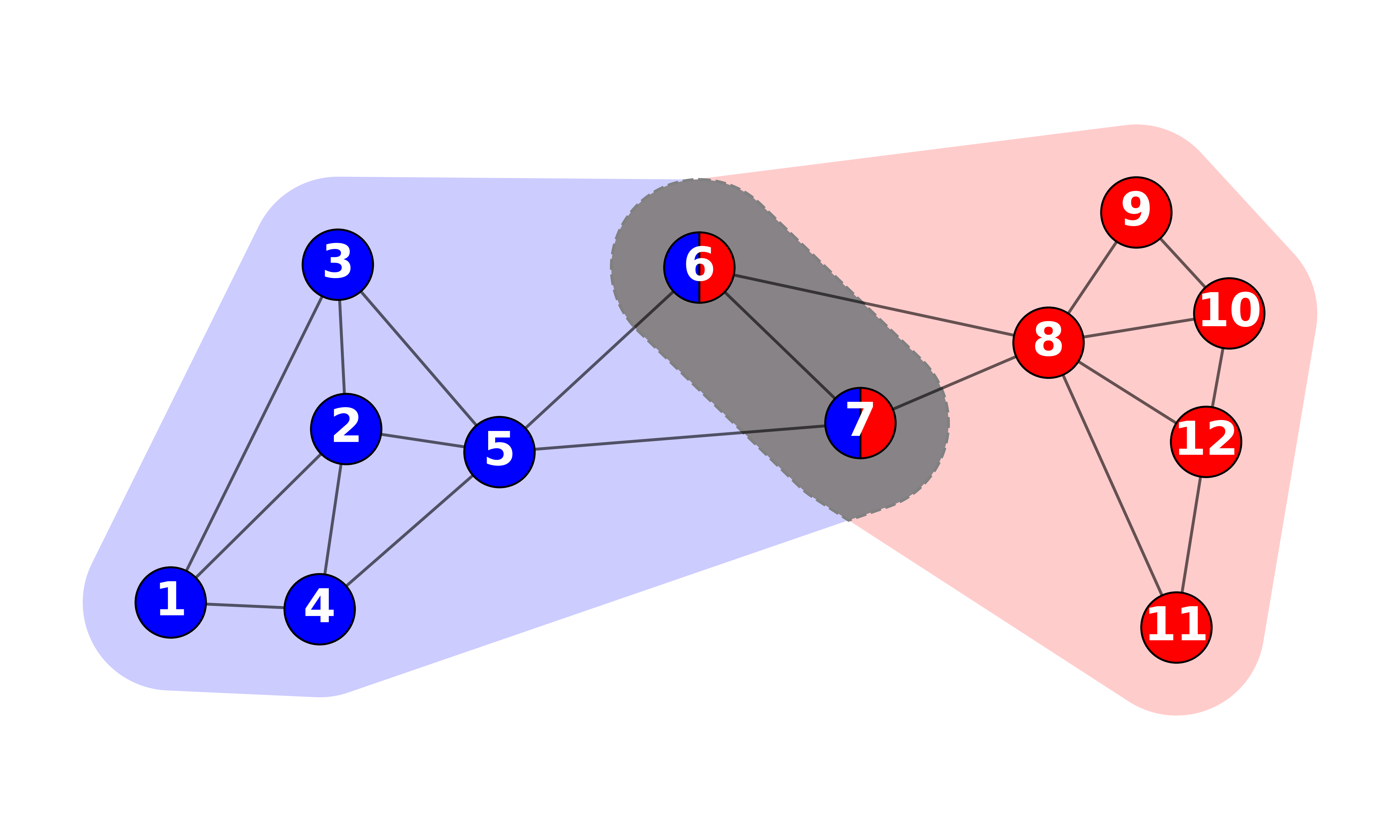}
    \caption{Overlapping Communities}
    \label{fig:illco}
  \end{subfigure}
  \caption{Illustration of a differences between circles and communities. A small network of 12 nodes with three circles (left) and two communities (right). Nodes 5 and 8 lie in Overlapping Circle regions, while the nodes 6 and 7 are in the intersection of two overlapping communities. %Overlapping community intersections.
  }
  \label{fig:illustrate}
\end{figure}

The distinction between communities and circles is often blurred in the literature, where these terms are sometimes used interchangeably \cite{shin2013study,brauer2014are}. In social network studies, an "overlapping node" often refers to a node present in multiple circles, although these studies have adopted the term \textit{community}\cite{yang2012structure_arxiv,palla2005uncovering}. This reflects the lack of a widely accepted definition of community structure. Although there is some correlation between the circles that overlap and communities \cite{shin2013study}, the latter are more a topological phenomenon, while the former are more local and context-specific and derived from the attributes of the nodes \cite{roy2024homophilic}. Topologically, circles may exhibit higher internal connectivity and a large number of external mediating links, while communities are cohesive internally but sparsely connected to the rest of the network \cite{brauer2014are}. In the present study, we focus on the overlapping nodes that participate in multiple circles and intentionally avoid the term "community". Although the function of circles vary across datasets (e.g., lists of friends, user-defined groups, or categorical labels), in all cases they are constructed from node attributes rather than inferred solely from network topology. Our aim is not to enforce a universal structural definition, but to investigate whether overlap across such attribute-defined substructures systematically affects spreading dynamics. The analysis of nodes belonging to multiple communities is left for future work. In Figure~\ref{fig:illustrate}, we illustrate the differences between circles and communities.

Overlapping nodes have been shown to play an important role in spreading processes in many real world settings, as they can act as bridges or hubs, and accelerate the spread of information or contagion \cite{shang2015epidemic,palla2005uncovering}. Neglecting the overlapping structure in spreading models can result in underestimating the reach and speed of propagation, as overlaps effectively create shortcuts that bypass the modular structure of the network \cite{rajeh2024role,peng2020network}. Recent studies have underscored the importance of incorporating overlap in spreading models to better capture the characteristics of real-world spreading processes \cite{kuikka2024detecting,modvit2022,chakraborty2016immunization}. Such processes include the spread of influence such as diseases, behaviour, opinion, information, or even cyberattacks through networks of various kinds.   
In these processes, we distinguish two types of mechanisms being either simple contagion (SC) or complex contagion (CC). In SC models, one assumes that information passes in a single node‐to‐node contact (e.g., Susceptible-Infected (SI) and Susceptible-Infected-Recovered (SIR) models).\cite{centola2007complex, almiala2023similarity} In contrast, in CC models, one allows reinforcement through multiple exposures, which captures better phenomena such as social reinforcement or cascading failures~\cite{blume2011}. Studying both SC and CC dynamics provides a deeper insight into how network topology and overlapping structure influence spreading behaviour in them, and provides a process-level control by comparing the spreading effects under two distinct contagion regimes.

In this study, we introduce a probabilistic framework for quantifying the role of overlapping nodes in spreading processes. Using centrality-based metrics, we analyse both the distributional and temporal characteristics of overlapping nodes in the propagation of influence or contagion. To describe the spreading process we apply both SC and CC models across multiple real-world network datasets. Within the studied datasets and influence spreading dynamics, our results indicate that nodes belonging to overlapping circles are associated with higher spreading influence compared to nodes confined to single circles. Importantly, the overlap is associated with increased spreading influence beyond what is explained by degree alone, although the two remain highly correlated. We further compare our findings with existing literature and evaluate the feasibility of using attribute-defined circles for analysis.
%
%
%
%
% END OF INTRO
%
%
%
%
%
%
%
%
% START OF RELATED WORK
%
%
%
\section*{Related Work}
Research on overlapping community structures has attracted  
attention since the early 2000s \cite{palla2005uncovering, lancichinetti2009detecting, gregory2009finding}. Although community overlap has been extensively studied, research focused specifically on differences in spreading processes remains limited. The distinct roles of overlapping and non-overlapping nodes of circles and communities have primarily been studied in the fields of epidemics and social sciences.

The differences in spreading between overlapping and non-overlapping nodes were investigated in \cite{ reid2011diffusion} and \cite{shang2015epidemic}, in which the analysis was performed by rewiring the networks and thus altering the network topology by creating more communities. The principle was that when adding inter-community edges, the overlapping nodes become bridges that bypass multi-step routes, i.e. the network becomes more integrated. For the spreading process the authors used both the CC and SC models. The overlapping nodes tend to play a key role when the nodes are rewiring. Similarly, in \cite{yang2012structure_arxiv}, the authors found that the most influential nodes in social networks are often at the intersections of multiple circles. Their empirical analysis with the real-world network datasets confirmed that the overlapping regions are typically more densely connected internally than the non-overlapping ones. This is consistent with overlaps emerging naturally in observed social structures rather than being artefacts of network manipulation, and that such nodes act as key propagators in real network dynamics.

In the case of epidemic modelling, SC models have been widely employed to capture the cascade-like spread of infections (see, e.g., \cite{pastor2015epidemic} for a comprehensive  survey). In \cite{shang2015epidemic}, the authors compared synthetic and real-world topologies by running simulations that alternately designate overlapping nodes as recovered or leave them susceptible, which conclusively shows that these intersectional nodes are the principal drivers of intensity and speed of the contagion outbreak. Finally, targeted immunisation strategies aim to pinpoint the structural importance of overlapping nodes to contain epidemics with minimal resources. As demonstrated in \cite{chakraborty2016immunization}, immunising overlapping nodes reduces the epidemic prevalence far more effectively than approaches focused exclusively on non-overlapping nodes, which once again underscores how overlap critically shapes both the propagation and control of contagion processes. We note that most of the studies discussed above focus on overlap defined primarily through network topology or community structure and not of attributes.

From a statistical perspective, the distribution of the centrality metrics of nodes has been explored recently in \cite{alasadi2024clustering}. The authors studied several different centrality metrics for nodes that reflect the node's importance in spreading processes. They discovered that the top quartile of a centrality metric contains more overlapping nodes contributing than non-overlapping. Many other works have similarly ranked spreading power by centrality (see \cite{Lawyer2015} and references therein), but this can be misleading. A high centrality score does not guarantee maximal spreading capability, and the total spreading power and node's centrality metrics are not necessarily analogous \cite{Lawyer2015}. In fact, nodes with only moderate centrality value can sometimes ignite large cascades and thus have a central role in diffusion processes \cite{cha2010measuring}. Centrality-aware metrics have also been suggested to find out the true power of a node (see, e.g. \cite{ghalmane2019centrality, zhao2015communitybasedcentrality, modvit2022}). These centrality-aware metrics, despite considering both local and global node properties, still rely heavily on structural information alone, without adequately accounting for the probabilistic and temporal nature of spreading processes. Thus, they might fail to capture the actual spreading dynamics, making them similarly unfit for accurately identifying the real sources of influence and truly important nodes. In the following sections, our aim is to address these issues with a probabilistic approach based on the influence spreading model.
%
%
%
%
% END OF RELATED WORK
%
%
%
%
%
%
%
%
% START OF METHODS AND DATA
%
%
%
%
%
%
%
%
\section*{Methods and Data}
\subsection*{Methods}
For analysis, the nodes are categorised into two groups, i.e., Overlapping nodes (OL): Nodes belonging to two or more circles and Non-overlapping nodes (NOL): Nodes belonging to fewer than two circles. We employ the probabilistic Influence Spreading Model introduced in \cite{kuikka2018influence}, a unified framework capable of modelling both SC and CC processes. The model outputs an Influence Spreading Matrix (ISM), denoted by $\textbf{C}$, where each entry $C_{ij}$ represents the probability that the influence, originating from node $i$, reaches node $j$. Thus, the model captures all pairwise interactions across the network. The SC model allows the influence to propagate only through self-avoiding paths, as in classical SI/SIR epidemic models. The CC model incorporates recurrent interactions and feedback loops, capturing threshold-like and higher-order effects that are characteristic of social reinforcement in real-world networks \cite{kuikka2023opinion}. For a full description of the model, see the definition in \cite{kuikka2018influence}. From the ISM, we calculate the following centrality metrics at each timestep $T$.
\subsubsection*{In- and Out-Centrality}
\textit{In-centrality} $\textbf{C}^{\mathrm{(in)}}$ is defined as the column sum of the ISM, reflecting a node's susceptibility or likelihood to receive influence from others:
%eq1
\begin{align} C^{(\textrm{in})}_{(j)}(T) \ = \sum _{\begin{array}{c} j \ne i \end{array}}C(i,j)(T) \ . \end{align}
%eq1
\textit{Out-centrality} $\textbf{C}^{(\textrm{out})}$ is given by the sum of the rows of the ISM, measuring the potential of a node to spread influence to other nodes throughout the network:
%eq2
\begin{align} C^{(\textrm{out})}_{(i)} \ = \sum _{\begin{array}{c} i \ne j \end{array}}C(i,j) (T) \ . 
\end{align}
%eq2
We calculate the relative difference between the OL and NOL nodes for each network $V$ and then aggregate these differences across all $N$ networks (where $N$ is the total number of networks), for both $\textbf{C}^{\mathrm{in}}$ and $\textbf{C}^{\mathrm{out}}$ at each time step $T$:
%eq3
\begin{align}
    \%\Delta_{\textbf{C}^{\mathrm{in|out}}}(T) &= 100 \cdot \frac{1}{N} \sum_{V=1}^{N} \times
    \frac{\langle C^{\mathrm{in|out}}_V(T) \rangle_{\mathrm{OL}} - \langle C^{\mathrm{in|out}}_V(T) \rangle_{\mathrm{NOL}}}{\langle C^{\mathrm{in|out}}_V(T) \rangle_{\mathrm{NOL}}} \ .
\end{align}
%eq4
\subsubsection*{Betweenness Centrality}
In addition, we calculate the third metric, namely the \textit{Betweenness Centrality}, derived from the ISM (as presented in \cite{kuikka2024detailed}). It aligns between the In- and Out-Centrality, as it reflects the mediating property of a node. The present metric differs
from traditional shortest-path-based approaches, e.g., those introduced in \cite{freeman1977measures,barabasi2016network}. Instead of relying solely on the shortest paths, the ISM allows one to consider all possible paths, which enables a more comprehensive assessment of the intermediary role of nodes in the network to be investigated. In \cite{modvit2022} and \cite{kuikka2024detailed}, for example, it was pointed out that traditional centrality measures do not capture the full influence of the nodes. Any traditional calculation of logic based on the shortest path might lead to an underestimation of the power of the nodes. 

The Betweenness Centrality $b_i$ of a node $i$ is defined based on the concept of network cohesion %. Cohesion 
$\mathscr{C}$ that represents the total influence across the entire network and is calculated as follows:
\begin{align}
\mathscr{C} = \sum_{\substack{i,j \in V \\ i \neq j}} C(i,j)
\end{align}
When node $i$ is removed from the network, the cohesion becomes
\begin{align}
\mathscr{C}_{i} &= \sum_{\substack{i,j \in V\setminus\{i\}\\i \neq j}} C(i,j).
\end{align}
Then the betweenness centrality $b_i$ is %then 
the relative decrease in cohesion due to the removal of node $i$:
\begin{align}
    b_{i} &= \frac{\mathscr{C} - \mathscr{C}_{i}}{\mathscr{C}} .
\end{align}

Now, the average Betweenness Centrality BC for %both 
OL and NOL nodes in $V$ with the subset size $s$ calculated over $T$, and the relative difference $\%\Delta_{\mathrm{BC}}$ between node classes across the networks is calculated as %with
\begin{align}
    \mathrm{BC}_V &= \frac{1}{s}\sum_{i \in \{\mathrm{OL|NOL}\}}  b_i \ ,
\end{align}

\begin{align}
    \%\Delta_{\mathrm{BC}}(T) &= 100 \cdot \frac{1}{N} \sum_{V=1}^{N} \times
    \frac{\langle \mathrm{BC}_V(T) \rangle_{\mathrm{OL}} - \langle \mathrm{BC}_V(T) \rangle_{\mathrm{NOL}}}{\langle \mathrm{BC}_V(T) \rangle_{\mathrm{NOL}}} \ .
\
\end{align}

\subsubsection*{Ratio Of Geometric Means}
To back up our observations, we also 
calculate the ratio of geometric means for saturated networks, i.e., when the spreading process has reached a steady state and no further propagation occurs. The arithmetic means are not sufficient as highly skewed data might introduce some bias. The metrics ($\textbf{C}^{\mathrm{(in|out)}}, \mathrm{BC}$), denoted here as $x$, often span several orders of magnitude, and the calculation of geometric means is less sensitive to large variations. We calculate the geometric means for the OL and NOL node classes as follows: %$G_{\mathrm{OL}}$ and $G_{\mathrm{NOL}}$ for both node classes:
\begin{align}
\mathrm{GM}_{\mathrm{OL}} &= 
 \Bigl(\prod_{\ \ i=1 , \ i\neq j }^{s} x_{i}\Bigr)^{\tfrac{1}{s}}
,\quad
\mathrm{GM}_{\mathrm{NOL}}
= \Bigl(\prod_{\ \ j=1, \ j\neq i}^{t} x_{j}\Bigr)^{\tfrac{1}{t}} \ ,
\end{align}
where $t$ and $s$ denote the sizes of the subset. We then calculate the ratio of geometric means $R$ given by 
\begin{align}
    R &= \frac{\mathrm{GM}_{\mathrm{OL}}}{\ \ \mathrm{GM}_{\mathrm{NOL}}} \ .
\label{eq:einstein}
\end{align}

\subsubsection*{Temporal Degree-Controlled Overlap Effect Analysis}

OL nodes typically have higher degree than their NOL counterparts. A natural subsequent question is whether their elevated in- and out-centrality is simply a consequence of degree, rather than of overlapping property itself. To assess whether overlap status exerts an effect beyond degree, we estimated node-level overlap effects separately for each network, diffusion time, and centrality metric. Specifically, for each node $i$ in network $V$ at time $T$, we modeled log-centrality as
\begin{align}
\log\!\big(C^{(M)}_{i,V,T}\big)
=
\alpha_{V,T,M}
+
\beta^{(M)}_{\mathrm{OL},V,T}\,\mathrm{OL}_{i}
+
\beta^{(M)}_{\mathrm{k},V,T}\,\log(1+k_i)
+
\eta_{i,V,T,M},
\label{eq:beta}
\end{align}

where \(C\) denotes the centrality value of node \(i\) for metric \(M\) in network \(V\) at diffusion time \(T\), \(M \in \{\mathrm{in},\mathrm{out}\}\), \(\mathrm{OL}_i \in \{0,1\}\) indicates overlap membership (0 for NOL and 1 for OL nodes), \(k_i\) is the degree of node \(i\), \(\alpha\) is the intercept, \(\beta_{\mathrm{OL}}\) is the overlap coefficient, \(\beta_{\mathrm{k}}\) is the degree coefficient, and \(\eta\) is the residual term. Thus, \(\beta_{\mathrm{OL}}\) represents the degree-controlled overlap effect. At each time $T$, we summarise network-level effects across networks using median. We finally convert the effect to percentages with
\begin{align}
\%\Delta_{\mathbf{C}^{(M)}} = 100\cdot(e^{\beta_{\mathrm{OL}}}-1),
\label{eq:betapercent}
\end{align}
which yields the degree-corrected relative difference in centralities between OL and NOL nodes.

To complement the descriptive summaries of the network-specific overlap coefficients, we use the Wilcoxon signed-rank test to assess whether the distribution of degree-controlled overlap effects across networks was systematically shifted above zero at each diffusion time. Specifically, for each time point \(T\) and metric \(M\), we consider the set of network-level coefficients
\begin{align}
\mathcal{B}_{T,M}
=
\left\{
\beta^{(M)}_{\mathrm{OL},V,T}
:\;
V=1,\dots,N_T
\right\},
\end{align}
where \(N_T\) denotes the number of networks contributing estimates at time \(T\). We then test for each dataset
\begin{align}
H_0 &: \operatorname{median}\!\left(\mathcal{B}_{T,M}\right)=0,\\
H_A &: \operatorname{median}\!\left(\mathcal{B}_{T,M}\right)>0,
\end{align}
using the Wilcoxon signed-rank statistic applied to the values
\(\beta^{(M)}_{\mathrm{OL},V,T}\).

\subsection*{Data}
Our analysis utilises ego-networks drawn from four sources: ego-Facebook \cite{mcauley2012learning} (FB), com-LiveJournal \cite{backstrom2006group,leskovec2009community} (LJ), com-Orkut \cite{nr} (ORK) and wiki-topcats \cite{yin2017local,klymko2014using} (wiki). The first three are undirected social networks, while the wiki-topcats dataset is originally a directed hyperlink network. Since our analysis focuses on circle membership and overlap rather than on direction-specific dynamics, we symmetrised the wiki network by treating each directed edge as an undirected connection. For clarity, we use the abbreviations in what follows.

The key characteristics of the networks and circles are summarised in Table \ref{tab:network_datasets}. We extracted each subnetwork by selecting nodes between 500 and 1 500 neighbors from the full graph. We deliberately chose datasets with diverse structural properties so that our results could capture universally common characteristics of nodes rather than reflecting the properties of highly similar networks. Circles were then constructed from the provided circle assignments. The fraction of OL nodes varies substantially: In some networks, only a small subset of nodes overlap, whereas in others the majority do. For the LJ and ORK datasets, we only considered circles that contain at least ten nodes in the network. The decision for this threshold is addressed in the Discussion section. Finally, to calculate the  
Betweenness Centrality, we analysed a reduced set of networks to limit computational costs: we included all four FB networks, and for the rest of the datasets, we used subsamples of 20 and 10 networks for the CC and SC models, respectively.

To mitigate the inherent bias associated with the ego node, we set its node probability to zero, effectively removing the central node and any resulting isolated components from the analysis. The rest of the nodes' weights are set to 1, and the edge weights are uniformly set to 0.05, which provides a balance between overly rapid saturation and negligible propagation. Additional sensitivity analysis with weights in the range $[0.001, 1.0]$ confirms that the qualitative differences between OL and NOL nodes remain consistent (see Appendix B). The maximum path length is set to 100 to account for far-reaching influence without excessive computational cost. The time parameter $T$ between 1 and 100 is used to capture the temporal dynamics of the spreading. See the full definitions of the parameters in \cite{kuikka2018influence,Kuikka_Aalto}.
\begin{table}[ht]
    \centering
    \resizebox{\textwidth}{!}{%
    \begin{tabular}{l c c c c c c c}
        \toprule
        Dataset name    & Abbr. & N & Nodes              & Clustering           & Avg.\ degree             & Overlapping attribute                  & Overlap\,\%       \\ 
        \midrule
        ego‑Facebook    & FB           & 4           & 760 (532–1034)      & 0.54 (0.47–0.63)     & 44.0 (18.1–80.8)        & Friends‑lists                          & 7.0 (1.1–34.4)    \\
        com‑LiveJournal & LJ           & 51          & 1176 (833–1486)     & 0.27 (0.07–0.48)     & 12.0 (3.14–62.2)        & User‑created groups                    & 79.2 (23.3–97.1)  \\
        com‑Orkut       & ORK          & 27          & 926 (801–1284)      & 0.23 (0.07–0.50)     & 11.4 (2.8–46.4)         & User‑created groups                    & 81.0 (42.0–94.9)  \\
        wiki‑Topcats    & WIKI         & 131         & 1127 (806–1495)     & 0.26 (0.15–0.49)     & 7.8 (3.0–23.2)          & Top 100 categories in Wikipedia         & 23.4 (1.2–94.7)   \\
        \bottomrule
    \end{tabular}%
    }
    \caption{The number of networks per dataset, Summary statistics (mean (min–max)) for network size, clustering coefficient, average degree, and overlapping attribute construction type and OL proportions across datasets.}
    \label{tab:network_datasets}
\end{table}
%
% end of data
%
%
%
% Start of Results
%
\section*{Results}
\subsection*{Distributions of Metrics}
\begin{figure}[ht]
  \centering
  % First row
  \begin{subfigure}[t]{0.48\linewidth}
    \centering
    \includegraphics[width=\linewidth]{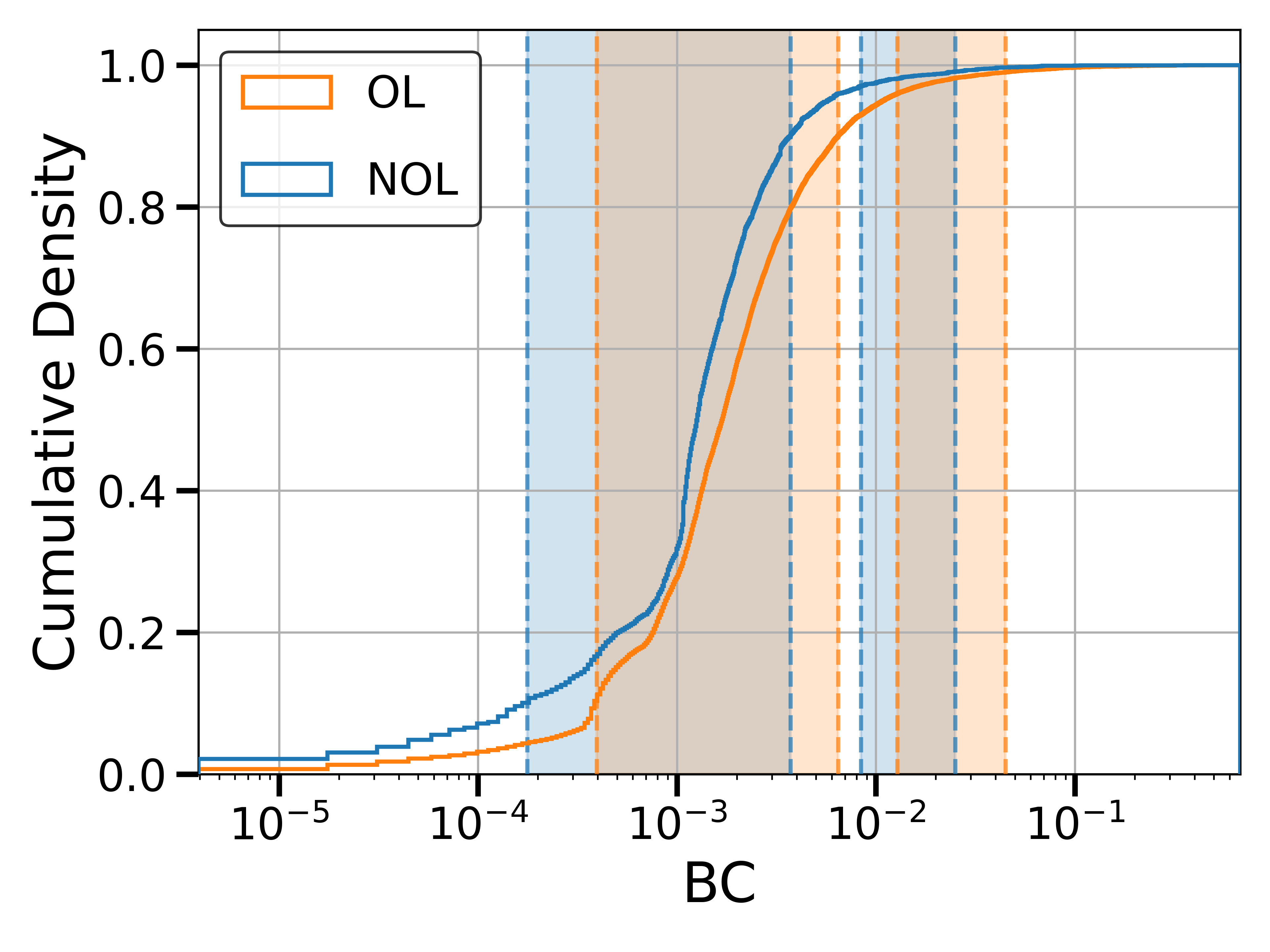}
    \caption{LJ}
    \label{fig:ccdf_LJ}
  \end{subfigure}
  \hfill
  \begin{subfigure}[t]{0.48\linewidth}
    \centering
    \includegraphics[width=\linewidth]{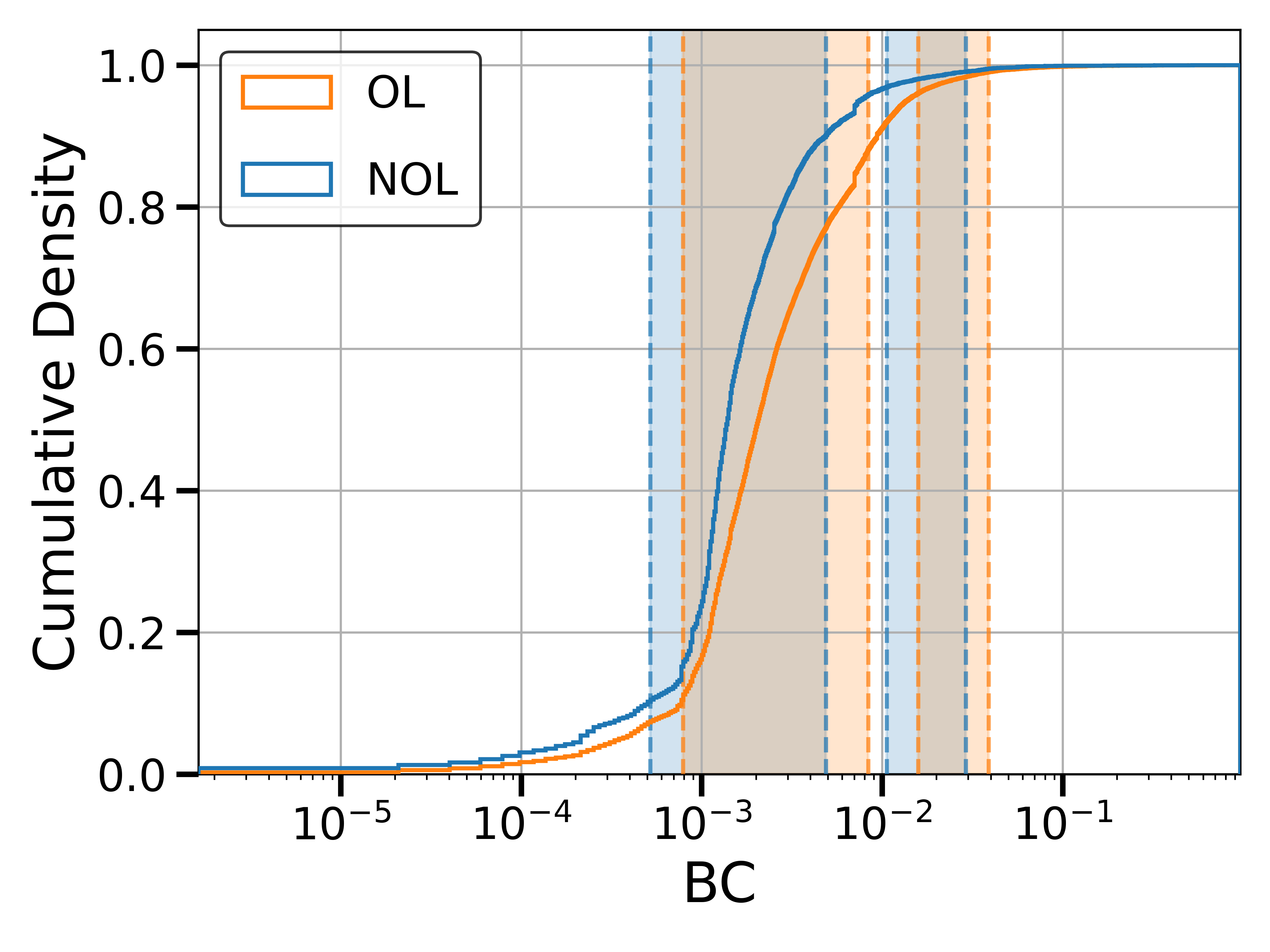}
    \caption{ORK}
    \label{fig:ccdf_ORK}
  \end{subfigure}

  \vspace{1em}

  % Second row
  \begin{subfigure}[t]{0.48\linewidth}
    \centering
    \includegraphics[width=\linewidth]{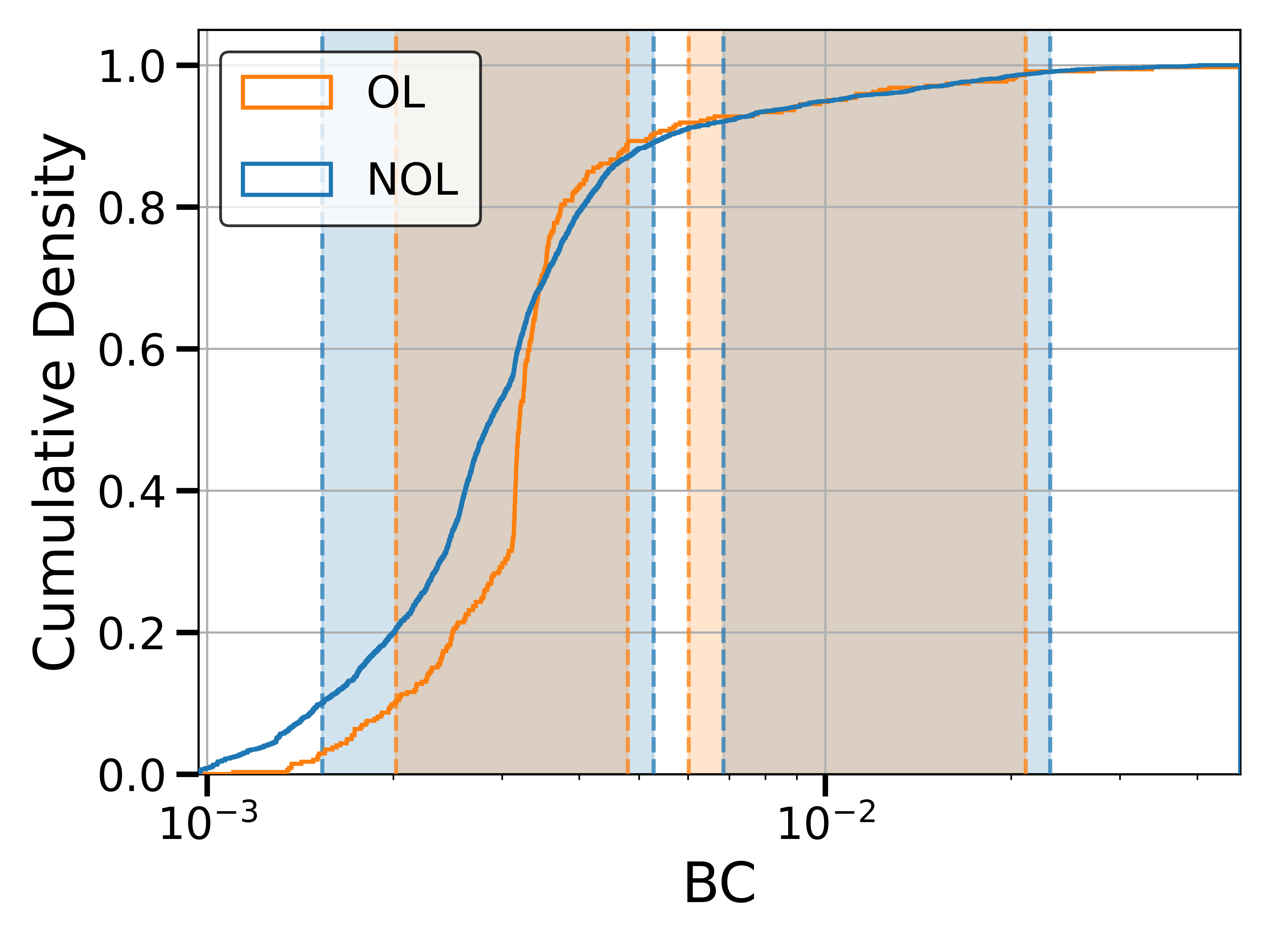}
    \caption{FB}
    \label{fig:ccdf_FB}
  \end{subfigure}
  \hfill
  \begin{subfigure}[t]{0.48\linewidth}
    \centering
    \includegraphics[width=\linewidth]{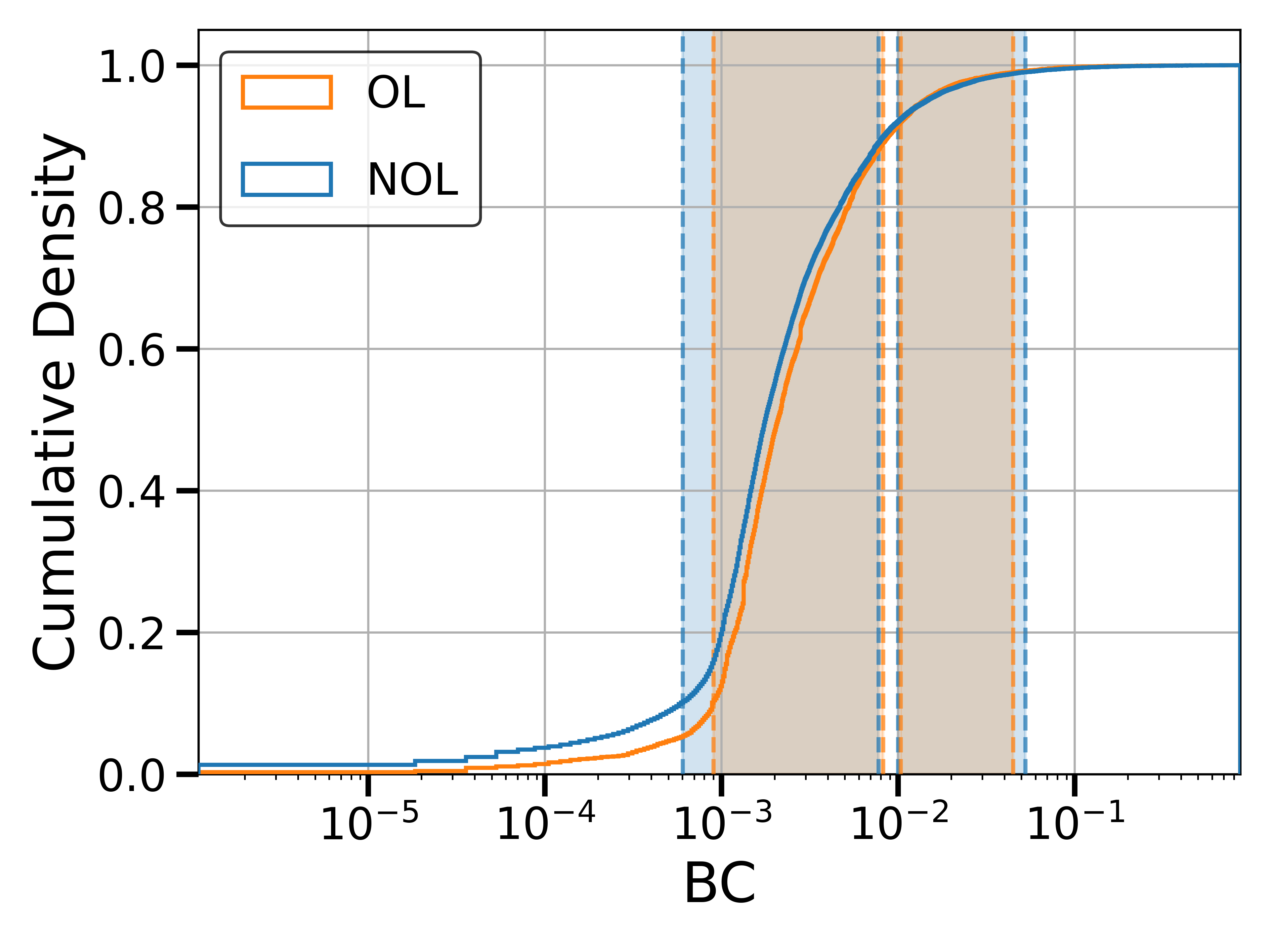}
    \caption{wiki}
    \label{fig:ccdf_wiki}
  \end{subfigure}

  \caption{Cumulative density of Betweenness Centrality (BC) of both OL and NOL nodes for Complex Contagion (CC) model. The first shaded areas from left represent the majority (80\%) of nodes, the latter the 91\%--99\% decile. A small amount of uniform jitter between classes has been added to distinguish these two percentile groups in the plot.}
  \label{fig:ccdfs}
\end{figure}

We started by first examining the properties of the individual nodes within a network to understand their contributions.  
We chose Betweenness Centrality as our metric due to its widespread use in the literature, and set the time parameter $T=30$. The metric was used, for example, in \cite{alasadi2024clustering} for a similar comparison, albeit with a different definition of Betweenness Centrality. We pooled all node-level metric values per dataset into a single aggregated distribution rather than treating each network separately. We analysed the distribution of node-level metrics in saturated networks for the CC model.  Figure \ref{fig:ccdfs} presents the cumulative distribution functions (cdf:s) for the sets of OL and NOL nodes, calculated for four different datasets. The first two shaded groups from the left (between the vertical dashed lines) represent the central 80\% of the OL and NOL distribution, bounded by the 10th and 90th percentiles. The second pair of shaded groups corresponds to the 91–99\% percentile range.

%\vspace{\baselineskip}
The distribution analysis indicates a notable shift in the bulk Betweenness Centrality values between OL and NOL nodes such that the bulk of NOL nodes generally lie left of the OL bulk, revealing that the NOL nodes typically possess lower Betweenness Centrality. A similar trend appears within the top 10\% of nodes in the case of LJ and ORK datasets. In contrast, the FB dataset exhibits an opposite shift, with OL nodes slightly displaced to lower centrality values, while the wiki dataset shows nearly equivalent distributions between the OL and NOL nodes within the top 10\% range. The FB dataset exhibits a different behaviour, where the OL nodes do not dominate in the upper tail. This may be partly explained by the low proportion of overlapping nodes, but also reflect the nature of the circle definition, i.e., the friendship lists may not capture structurally cohesive or topologically meaningful subgraphs in the same way as group-based or category-based circles. Consequently, the overlap in the FB networks may not correspond to structurally influential positions, which highlights a limitation of attribute-based circle definitions in the low-overlap settings.

\begin{figure}
    \centering
    \includegraphics[width=0.5\linewidth]{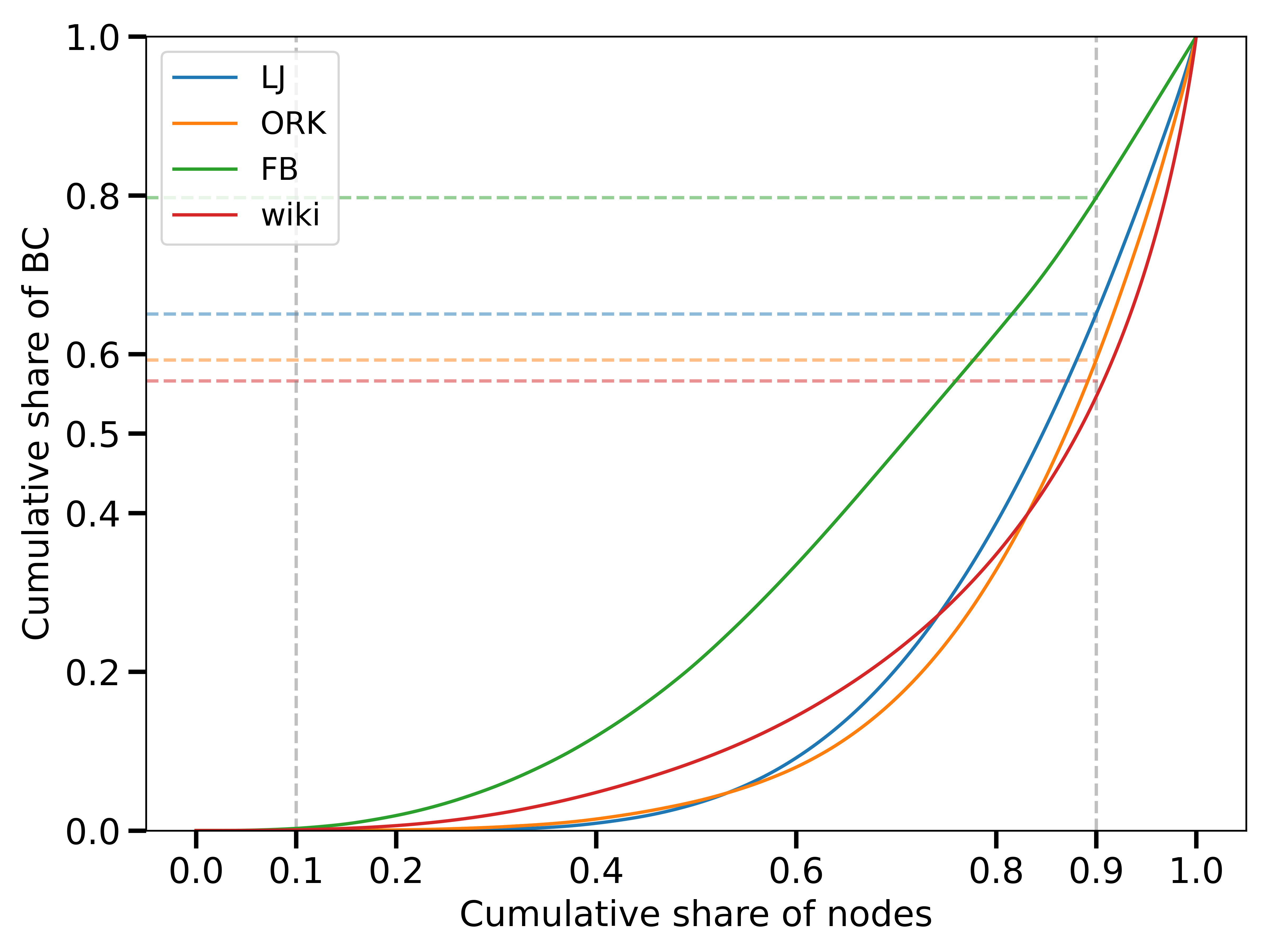}
     \caption{The Lorenz curves of Betweenness Centrality distributions. The dashed grey lines mark the 10th and 90th percentiles. The dashed coloured lines represent the proportion of Betweenness Centralities that fall in the bulk.}
    \label{fig:lorentz}
\end{figure}

The reason for choosing the specific percentile thresholds is further illustrated by the Lorenz curves presented in Figure \ref{fig:lorentz}. These curves elucidate how the Betweenness Centrality is unevenly distributed across nodes, highlighting the necessity to examine both bulk and extreme regions separately. In large datasets such as LJ, ORK, and wiki, the bulk nodes collectively contribute between 57–66\% of the total Betweenness Centrality. However, for the FB dataset, this contribution is even higher, highlighting the crucial role of the central bulk nodes. Nevertheless, the upper tail, particularly the top 10\%, still contributes substantially, i.e., 34--43\% for the LJ, ORK, and wiki datasets. This supports the considerable influence associated with high-centrality nodes. On the other hand the contribution of the bottom 10\% is negligible for all four analysed datasets.
Additionally, our inspection of the top 1\% highlights the prominent but varying role of superinfluencers, who alone account for approximately 8\% of the total share of Betweenness Centrality in the case of wiki networks, but with smaller contributions ($\lesssim5\%$) in the case of other three datasets. The upper tails of the Betweenness Centrality distributions approximate the power-law \cite{fairbanks2013statistical}, and therefore, to accurately address the imbalance between the OL and NOL groups, we have used exponential weighting for studying the top 1\% contributions between the OL and NOL groups. The results of the proportions are shown in Table \ref{tab:overlap_top_bottom}. Although the OL nodes are overpopulated in the ORK and LJ datasets for the decile and top 1\%, their contribution to the cumulative share is small. In contrast, the wiki dataset, albeit with a smaller proportion of OL nodes in both top 10\% and 1\%, is associated with a much larger share of Betweenness Centrality. 
\begin{table}[ht]
    \centering
    \begin{tabular}{l|c|c|c|c|c|c|c} \hline
        Dataset & Total& Top 10\% & Top 1\% & $N_{10\%}$ & $N_{1\%}$ & Lorenz$_{10\%}$ & Lorenz$_{1\%}$ \\ 
         &&&&&&& \\ \hline
        LJ   & 77.1 & 87.8 & 89.2 & 5913  & 591 & 35.9 & 4.18  \\
        ORK  & 79.3 & 88.2 & 92.9 & 2722 & 27 & 41.1 & 5.16 \\
        wiki & 27.3 & 30.3 & 31.2 & 14680 & 1468 & 43.8 & 8.00 \\
        FB   & 11.2 & 10.3 & 15.1 & 316  & 32 & 20.5 & 2.06 \\
    \end{tabular}
    \caption{Proportion of overlap (OL\%) nodes by dataset, in the weighed top 10 \% and top 1\% of Betweenness Centrality; the corresponding node counts ($N$); and each group’s share of total Betweenness Centrality (Lorenz).}
    \label{tab:overlap_top_bottom}
\end{table}

\subsection*{Temporal effects}
\subsubsection*{In- and Out-Centrality}
\begin{figure}[ht]
    \centering
    \begin{subfigure}[b]{0.48\linewidth}
        \centering
        \includegraphics[width=\linewidth]{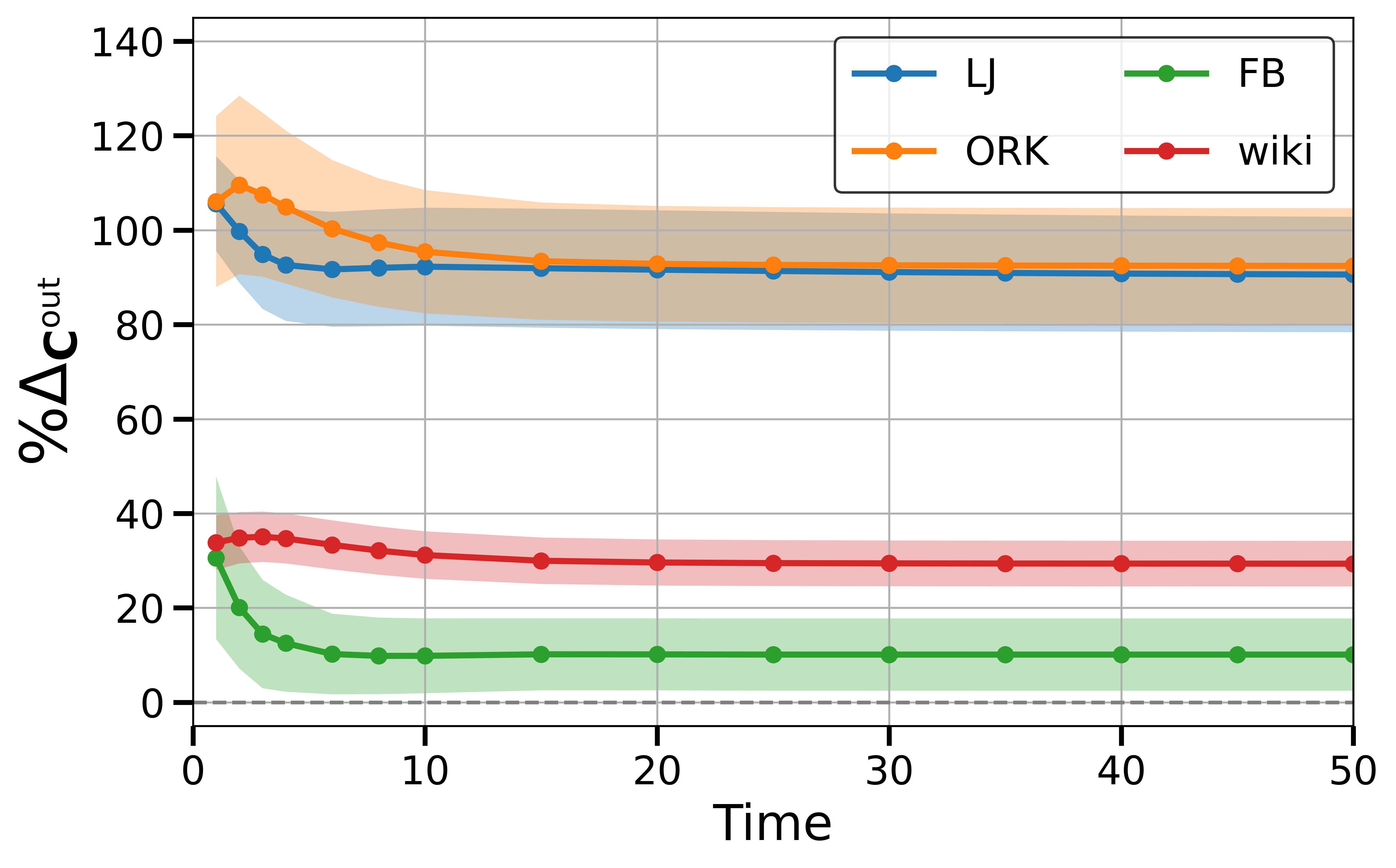}
        \caption{Out-Centrality}
        \label{fig:stream_a}
    \end{subfigure}
    \hfill
    \begin{subfigure}[b]{0.48\linewidth}
        \centering
        \includegraphics[width=\linewidth]{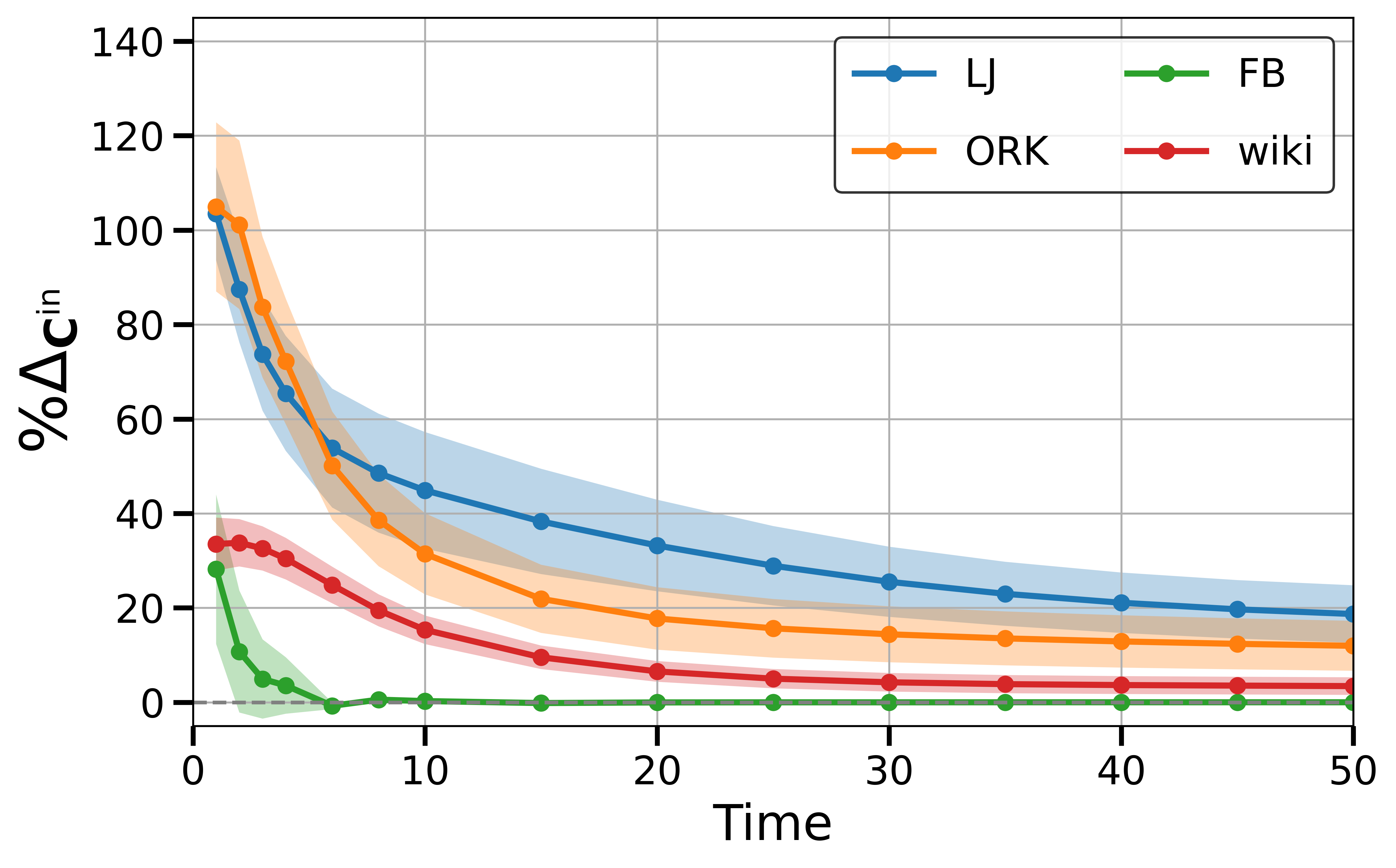}
        \caption{In-Centrality}
        \label{fig:stream_b}
    \end{subfigure}
    \caption{Complex Contagion: Comparison of OL and NOL nodes' relative difference. Out-Centrality (left) and In-Centrality (right) plotted with standard error mean (shaded).}
    \label{fig:stream_comparison}
\end{figure}

Next, we investigate how the spreading power between node classes evolves during the spreading processes. Figure \ref{fig:stream_comparison} illustrates the relative difference in the average In-centrality and Out-centrality between the OL and NOL nodes in the CC model, with the error representing one standard error of the mean. The OL nodes exhibit, on average, 90\% higher Out-centrality in the LJ and ORK datasets in the saturated phase, around $T\gtrsim15$; the wiki dataset shows 30\% higher Out-centrality, while the FB networks display a smaller difference. This is plausibly due to the comparatively smaller fraction of OL nodes in those networks. Nevertheless, the trend of decreasing and stabilising the Out-centrality in the beginning of spreading is visible in all datasets. Furthermore, the fluctuations observed in the early stages of spreading arise from the stochastic nature of path formation and the low initial probability of transmission. For small $T$, the influence propagates through a limited number of paths, making the dynamics highly sensitive to local topology of the network. As $T$ increases, the number of available paths grows, and the relative differences stabilise.
A common characteristic of the studied datasets are that they show continuous elevation of relative Out-Centrality. This persistence may be related to the bridge-like position of the OL nodes between multiple circles, which is consistent with their capacity to spread information. In contrast, the In-centrality gap between the OL and NOL nodes is most pronounced at the start of the spreading but narrows smoothly as spreading proceeds. This pattern suggests that the OL nodes are initially more susceptible to incoming contagion, 
due to their greater exposure, they also played an important role in the early spreading phase. At the end of spreading, the In-Centrality difference converges gradually and slower, much slower 
than the Out-Centrality, i.e., holding their susceptibility, although a small difference (1--20\%) of susceptibility for both node classes remain in most of the studied networks.

Finally, in wiki networks, an initial increase in both the In- and Out-Centrality measures, before their decline, reveals a short accumulation period. For In-Centrality, the delayed and gradual decline suggests a short incubation period, where the OL nodes retain their susceptibility. On the other hand, the increased relative difference of Out-Centrality describes another aspect of the process. It is delayed a bit more than In-Centrality because the spreading occurs more likely through the OL nodes than the NOL nodes. After accumulation, the OL nodes reach peak relative influence, after which the spread becomes slower. This threshold is 
not clearly visible and may depend on the balance between clustering and the average degree \cite{Gleeson2010}. The low average degree and low edge weights prevent the threshold from being reached immediately.

\begin{figure}[ht]
    \centering
    \includegraphics[width=0.48\linewidth]{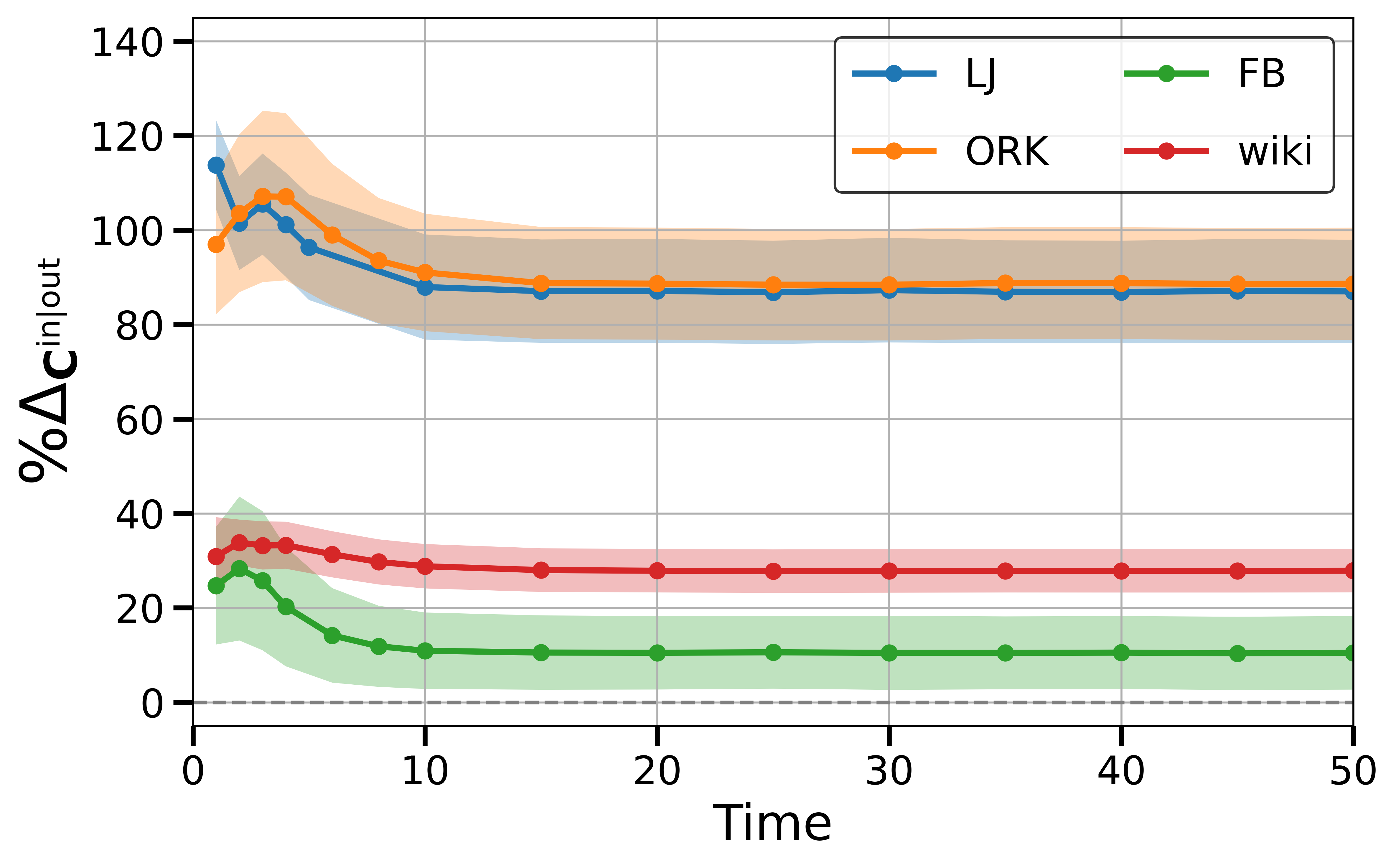}
    \caption{Simple Contagion: The network's relative In- and Out-Centrality differences with standard error means.}
    \label{fig:all_sc}
\end{figure}

Self-avoiding paths yield equivalent results for both In-Centrality and Out-Centrality and, therefore, Figure \ref{fig:all_sc} only presents the Out-Centrality results across all four datasets. In undirected graphs, every self-avoiding path from a source node to a target node corresponds to a reverse path from the target node back to the source node. However, in directed graphs or with more complex interactions that are present in the spreading process, the Out-Centrality and In-Centrality do not necessarily have equal values. For example, in our Complex Contagion model, reinforcement caused by cyclic and recurrent propagation breaks this symmetry.~\cite{kuikka2024detailed}
One observation is the similarity in the relative differences between the Out-Centrality of CC (Figure \ref{fig:stream_a}) and the Centralities of SC (Figure \ref{fig:all_sc}). The main difference is the presence of accumulation periods at the beginning of the SC spreading for the FB and LJ datasets, which are absent
in the CC case. Otherwise, the models show only minor variations within the standard error of the mean. Two factors may contribute to this similarity: First, the CC model includes a temporal delay at the beginning of the spreading process, requiring three propagation steps before reinforcement. This delay leads to negligible accumulation, making the spreading dynamics in both SC and CC models initially quite similar in the early stages. The choice of weights balances model differentiation with the ability to resolve spreading dynamics. Second, the relative differences between the OL and NOL nodes remain stable in the saturated regime, since our SC model follows SI dynamics. As a result, the high relative difference in the spreading potential tends to persist in saturated networks, similar to the CC model.
%Start of betweenness centrality
\subsubsection*{Betweenness  Centrality}
\begin{figure}[ht]
\centering
% Top row: CC and SC
\begin{subfigure}[t]{0.48\linewidth}
\centering
\includegraphics[width=\linewidth]{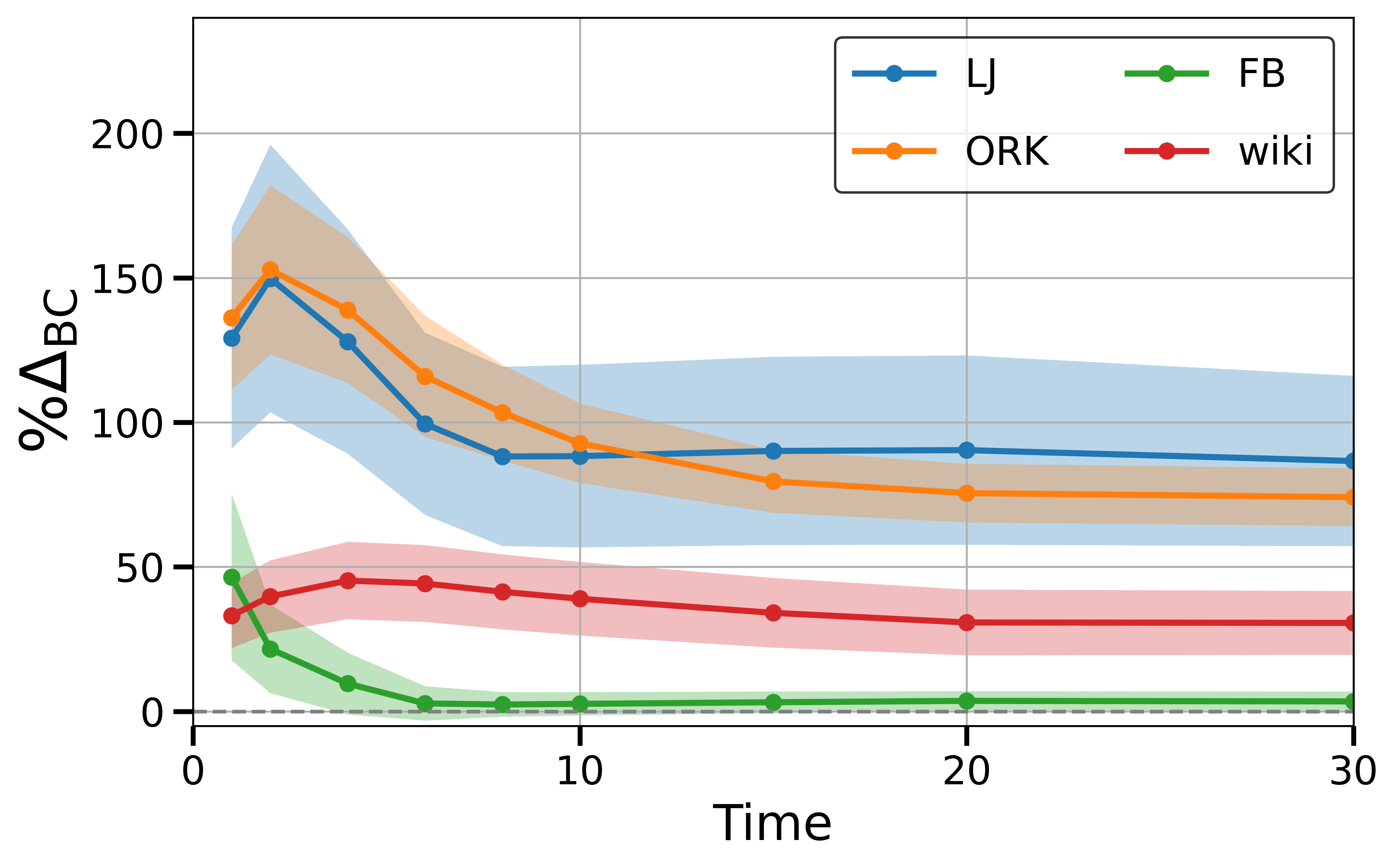}
\caption{(CC)}
\label{fig:betcc}
\end{subfigure}
\hfill
\begin{subfigure}[t]{0.48\linewidth}
\centering
\includegraphics[width=\linewidth]{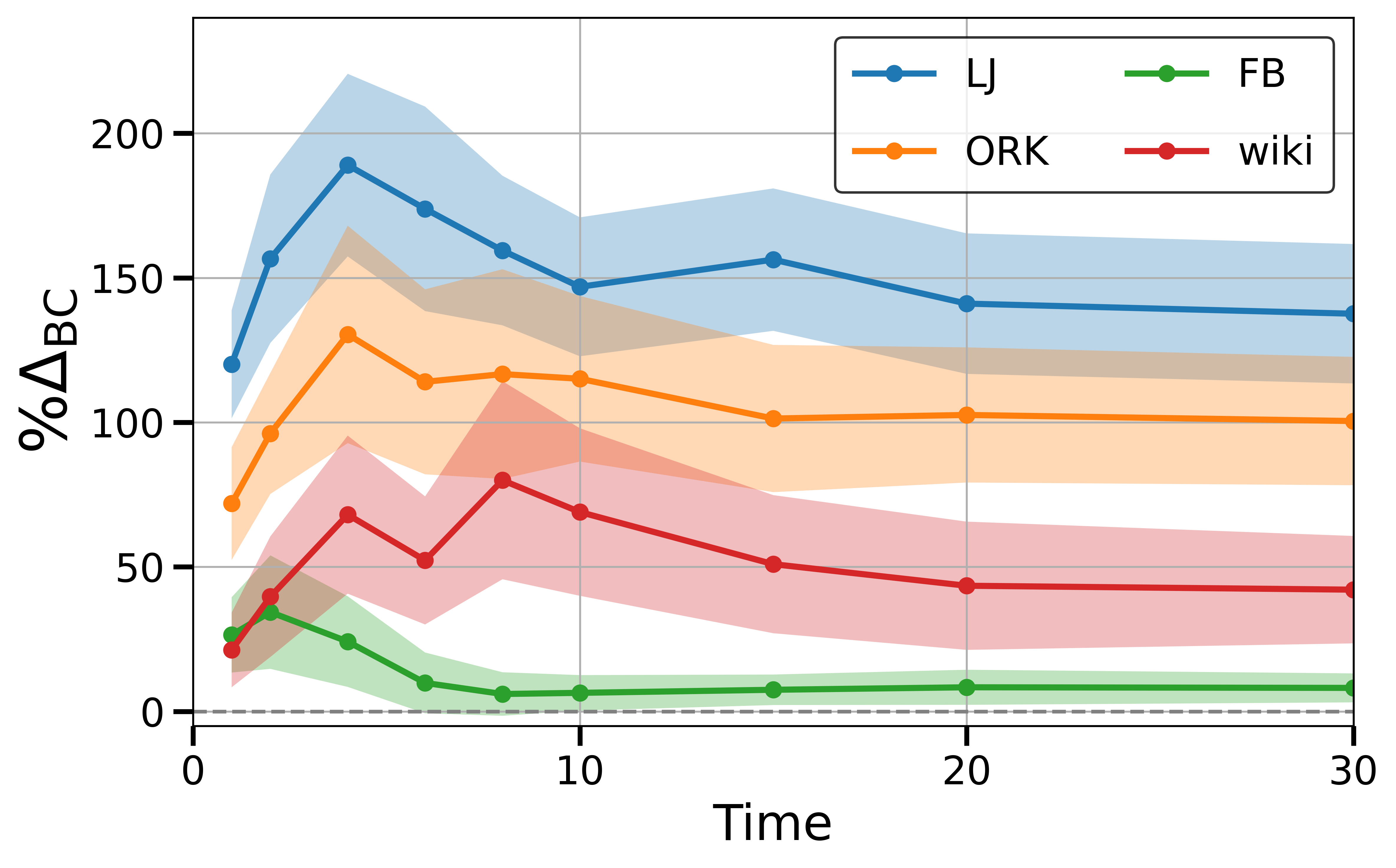}
\caption{(SC)}
\label{fig:betsc}
\end{subfigure}
  \caption{Temporal evolution of relative Betweenness Centrality for Complex Contagion (left) and Simple Contagion (right) models. Simulations were run on a subsample of networks: all FB networks were included, while for the remaining datasets we randomly sampled 20 networks for Complex Contagion and 10 for Simple Contagion.}
  \label{fig:bet}
\end{figure}

The Betweenness Centrality results for both Complex Contagion and Simple Contagion model simulations are shown in Figure \ref{fig:bet}. We observe a similar evolution of relative differences, although the SC curve is delayed. In the early stages of the process—before saturation begins—there is an accumulation period for the ORK, wiki, and LJ networks, which ends roughly before $T<5$ and $T<8$ for CC and SC, respectively. As nodes receive more exposures, their betweenness increases. OL nodes tend to receive more exposures, so their relative importance remains higher than that of NOL nodes. The shift to the right in the SC maxima, compared to the CC, reflects the slower spread of influence in both the OL and the NOL cases. In the FB dataset under CC, this effect is not evident—likely due to its high average degree, clustering, and low OL\%, but the other networks show that nodes belonging to multiple circles remain comparatively influential even in later stages of the spreading process, even in saturated conditions. In particular, the relative difference curves for Betweenness Centrality (Figure \ref{fig:betcc}) and Out-Centrality (Figure \ref{fig:all_sc}) exhibit similar trends. Since computing Betweenness Centrality is computationally intensive, Out-Centrality may serve as a computationally efficient proxy in similar settings in future studies, though we emphasise that these plots represent relative differences, not the raw metrics. The similarity between Betweenness Centrality and Out-Centrality was previously observed in \cite{kuikka2024detailed}, and a deeper comparison of these metrics will be the subject of our future work.
%end of BC
%
\subsection*{Ratio of Geometric Means}
\begin{figure}[ht]
  \centering
  \begin{subfigure}[t]{0.48\linewidth}
    \centering
    \includegraphics[width=\linewidth]{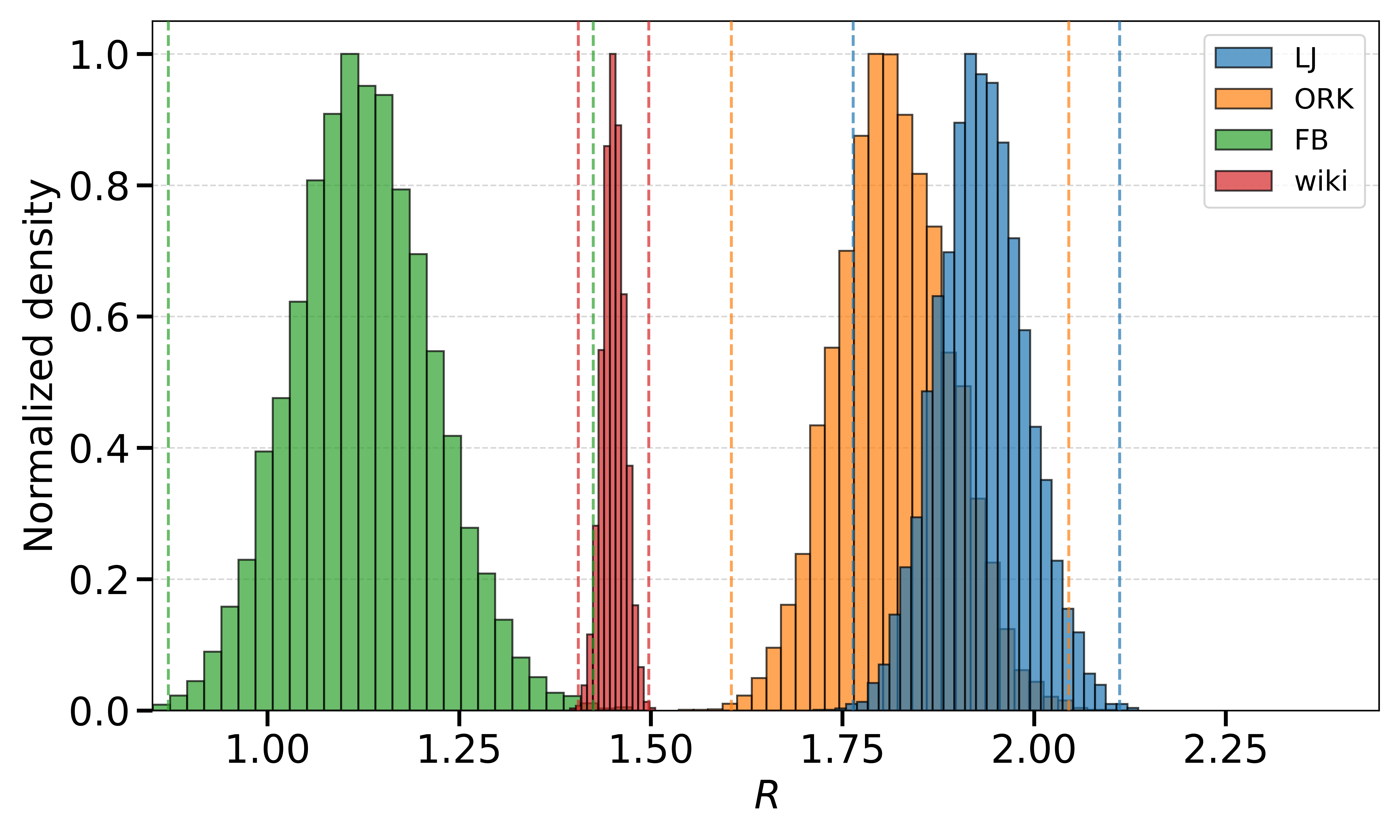}
    \caption{(CC): Out‑centrality}
    \label{fig:cc_out}
  \end{subfigure}
  \begin{subfigure}[t]{0.48\linewidth}
    \centering
    \includegraphics[width=\linewidth]{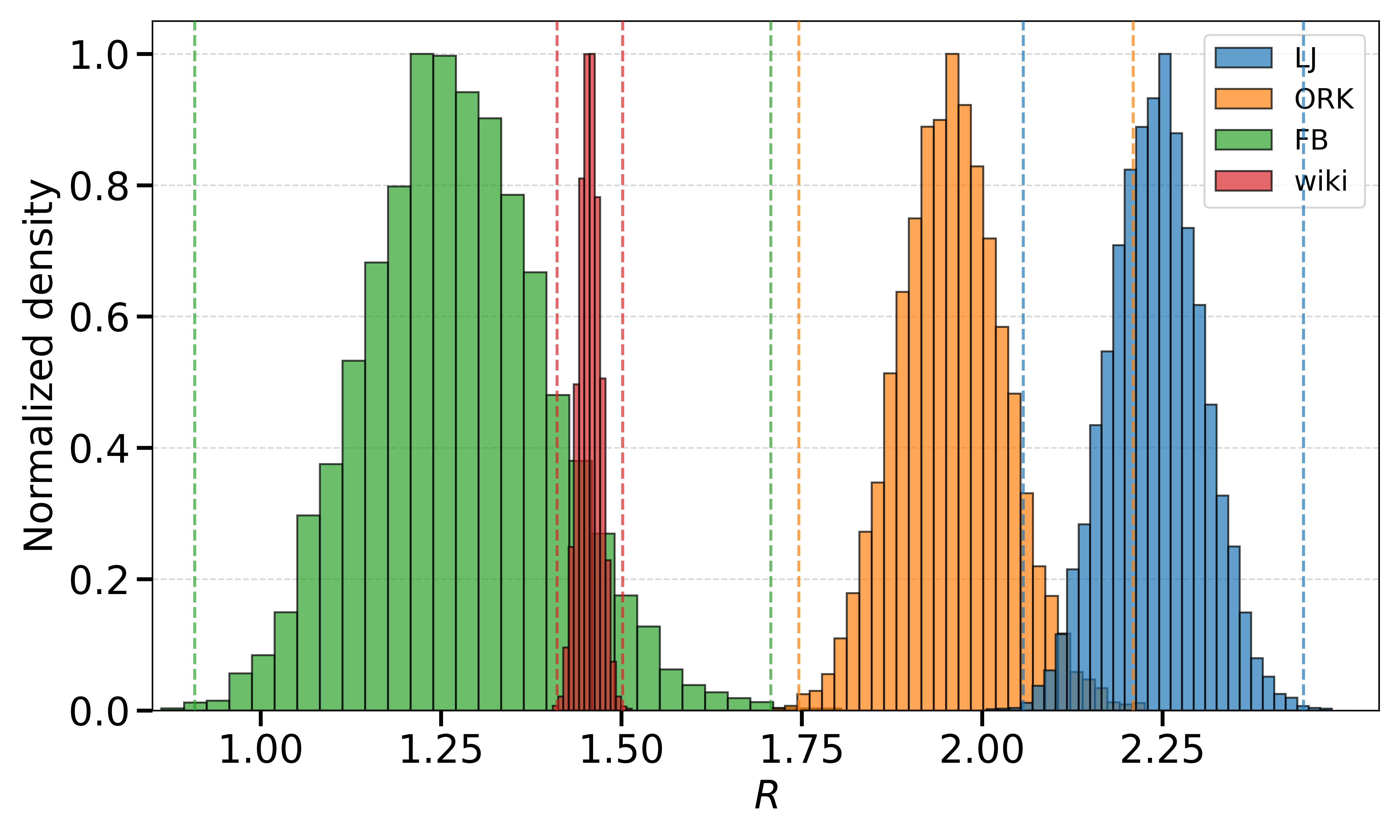}
    \caption{(SC): In- and Out‑centrality}
    \label{fig:sc_out}
  \end{subfigure} \hfill
   \vspace{1em}
   \begin{subfigure}[t]{0.48\linewidth}
    \centering
    \includegraphics[width=\linewidth]{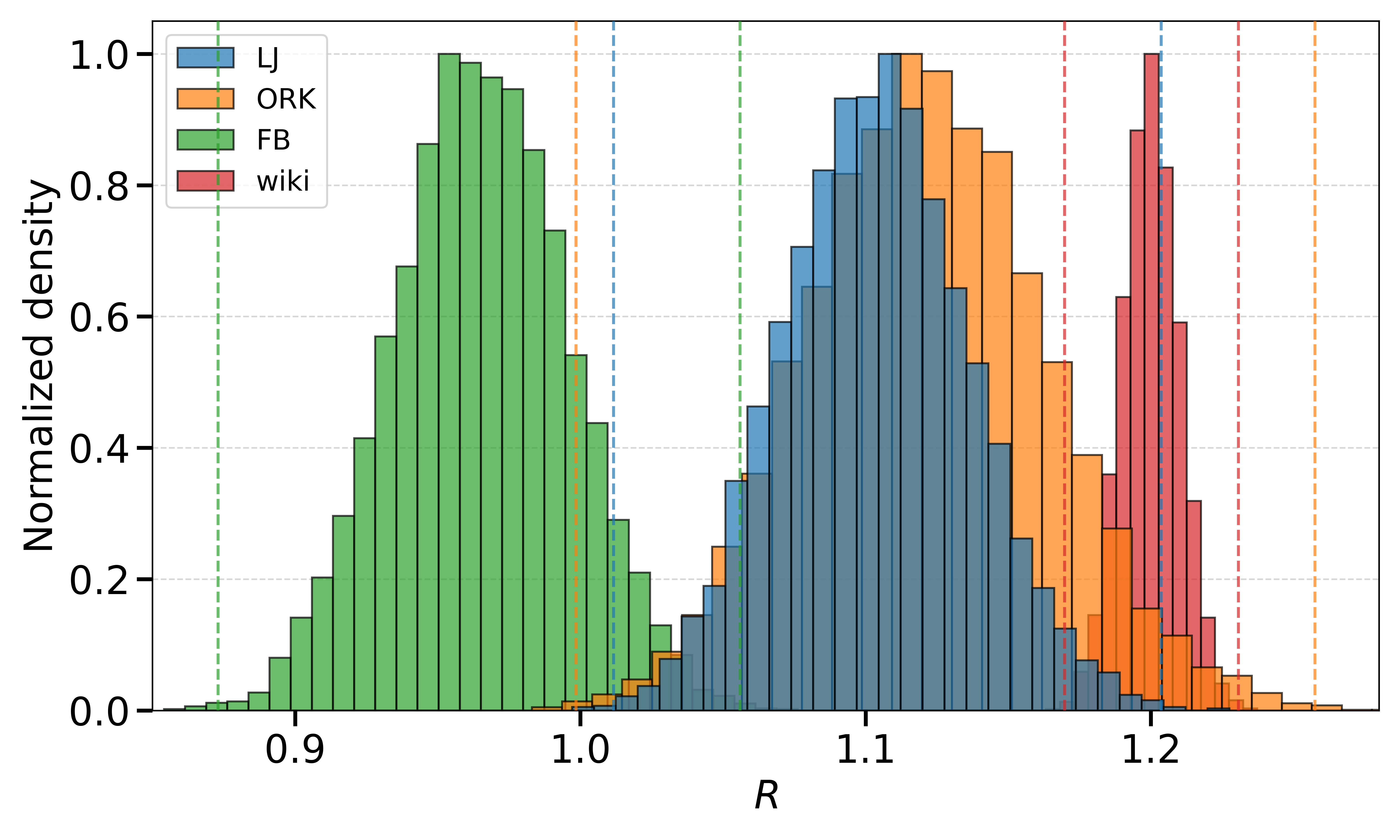}
    \caption{(CC): In‑centrality}
    \label{fig:cc_in}
  \end{subfigure}\hfill

  \caption{Geometric mean–ratio distributions via bootstrapping for $T=50$.}
  \label{fig:three_metrics}
\end{figure}

\begin{figure}[ht]
  \centering
  \begin{subfigure}[t]{0.48\linewidth}
    \centering
    \includegraphics[width=\linewidth]{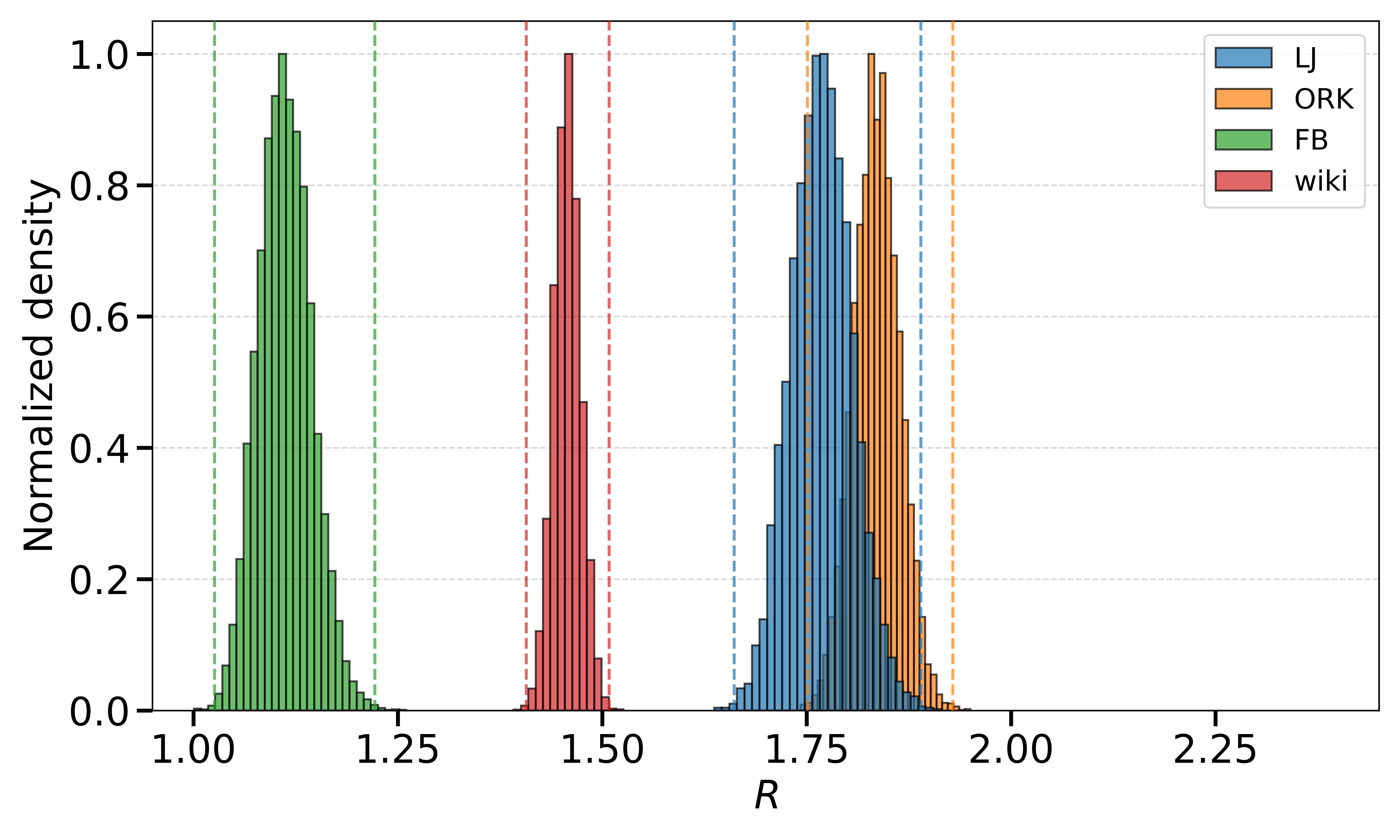}
    \caption{(CC)}
    \label{fig:cc_betw}
  \end{subfigure}\hfill
  \begin{subfigure}[t]{0.48\linewidth}
    \centering
    \includegraphics[width=\linewidth]{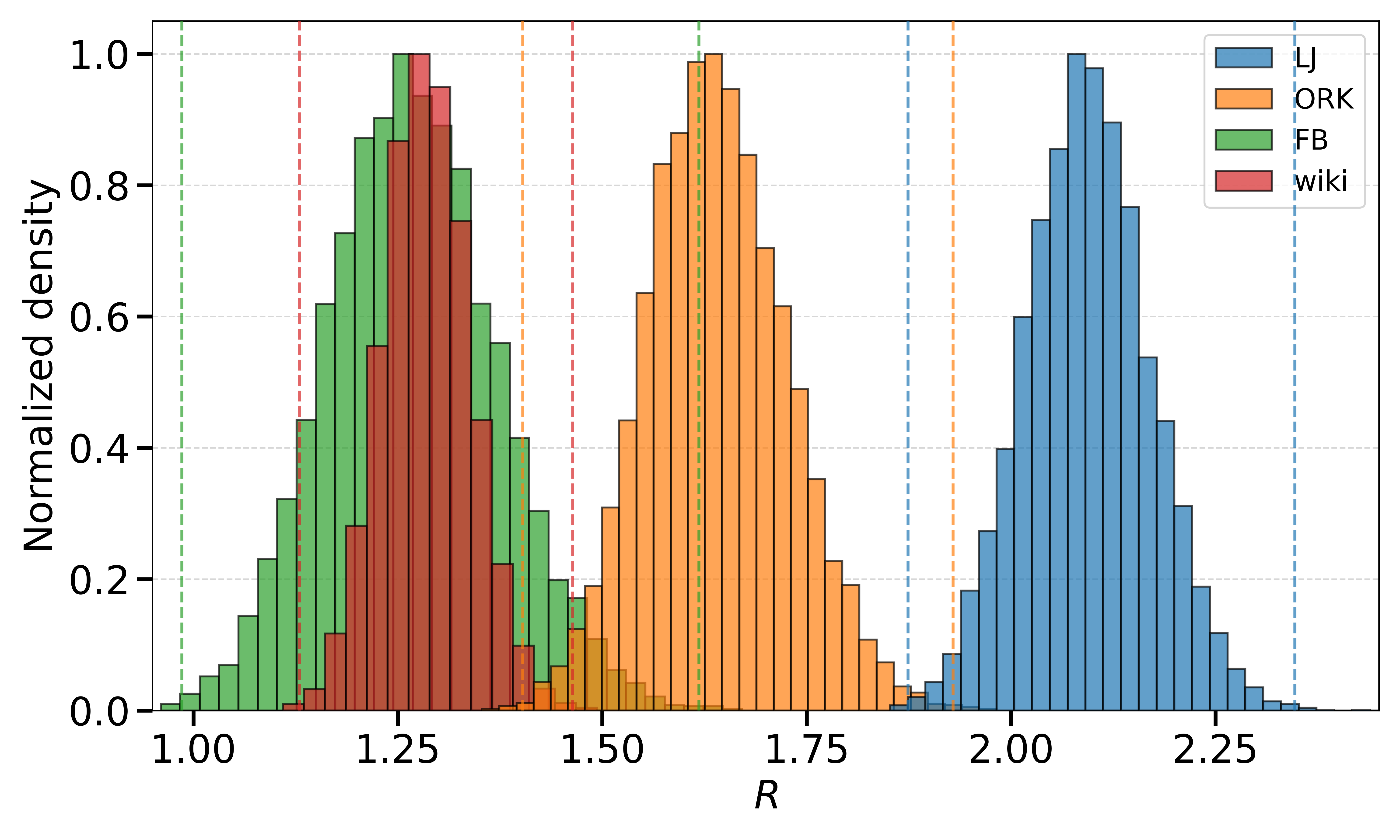}
    \caption{(SC):}
    \label{fig:sc_betw}
  \end{subfigure}
  \caption{Geometric mean–ratio distributions via bootstrapping for $T=30$ of Betweenness Centrality.}
  \label{fig:betw_metrics}
\end{figure}

To guard against bias in our relative comparisons, we repeated the geometric‐mean ratio analysis on saturated networks setting the time parameter threshold $T=50$ for In‑ and Out‑Centrality and $T=30$ for Betweenness Centrality. Given that the distributions of the metrics are highly skewed, we avoid relying on parametric assumptions and instead use a non-parametric bootstrap approach. Using a bootstrap with 10,000 resamples, we obtained the R‑ratios (Eq. \ref{eq:einstein}) and 1\%–99\% confidence limits, as depicted in Figures \ref{fig:three_metrics} and \ref{fig:betw_metrics}. For the four studied network datasets and for the CC and SC models, the R‑ratio for Out‑Centrality and Betweenness Centrality exceeds one, which indicates that OL nodes tend to have higher spreading properties than NOL nodes. The In‑Centrality R‑ratio concentrates around unity. The result is consistent with the decline towards one shown in Figure \ref{fig:stream_b}. Although FB shows greater sampling variability (and hence a slightly ambiguous R‑ratio), all previously observed relative results lie within the 99\% confidence limits. We also note the model‐dependent shifts such that the LJ, FB and ORK datasets show shift to the right under SC, while the wiki dataset shows no shift. As these shifts remain within the confidence limits, we refrain from drawing further conclusions. In general, the numeric ratios are consistent with the trends observed in the temporal analysis.

% Figure 2: CC and SC Betweenness

\subsection*{Degree-Contolled Overlap Importance}
\begin{figure}[ht]
\centering

\begin{subfigure}[t]{0.48\linewidth}
    \centering
    \includegraphics[width=\linewidth]{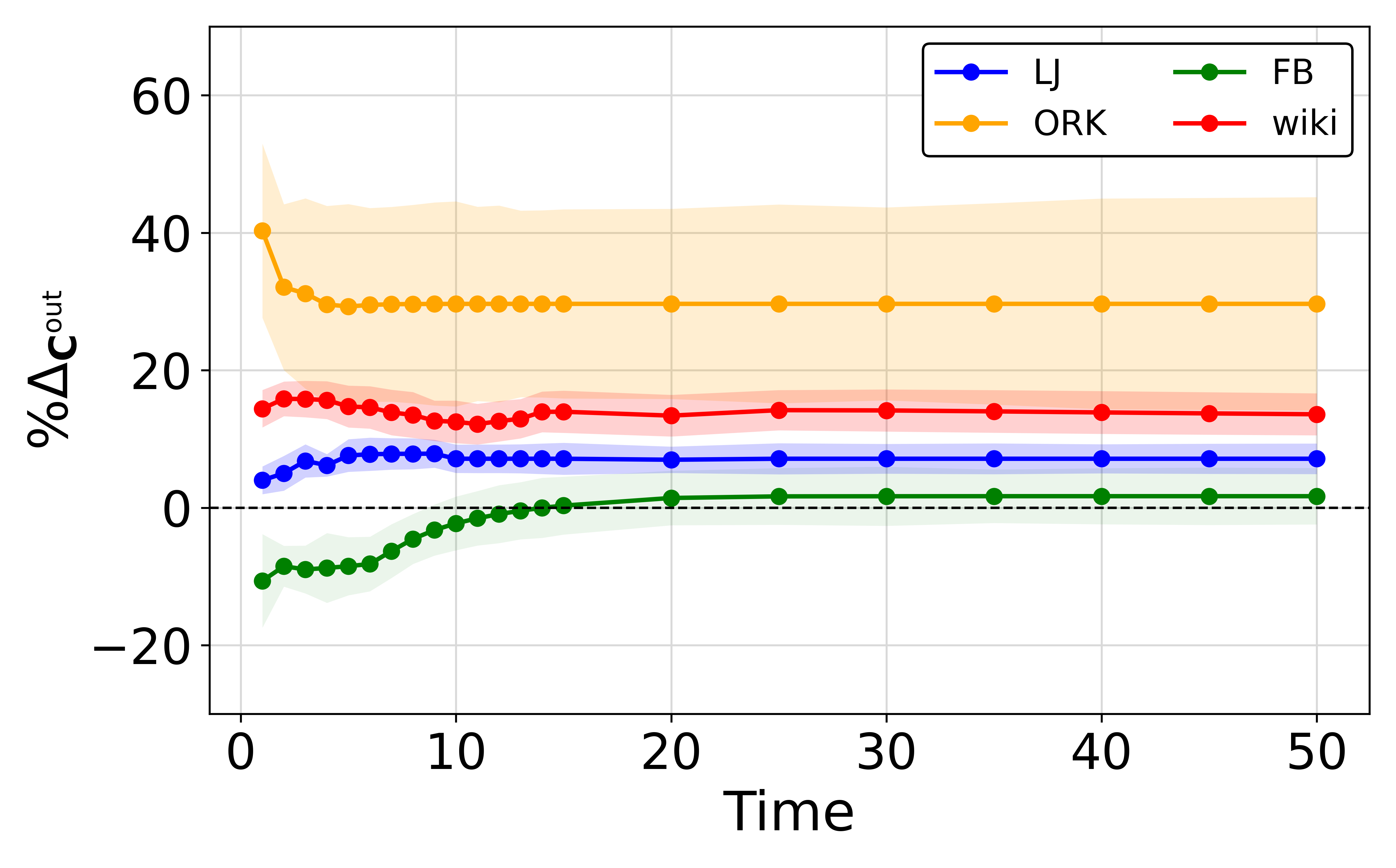}
    \caption{CC: Out-Centrality}
    \label{fig:NEW-CC-OC}
\end{subfigure}
\hfill
\begin{subfigure}[t]{0.48\linewidth}
    \centering
    \includegraphics[width=\linewidth]{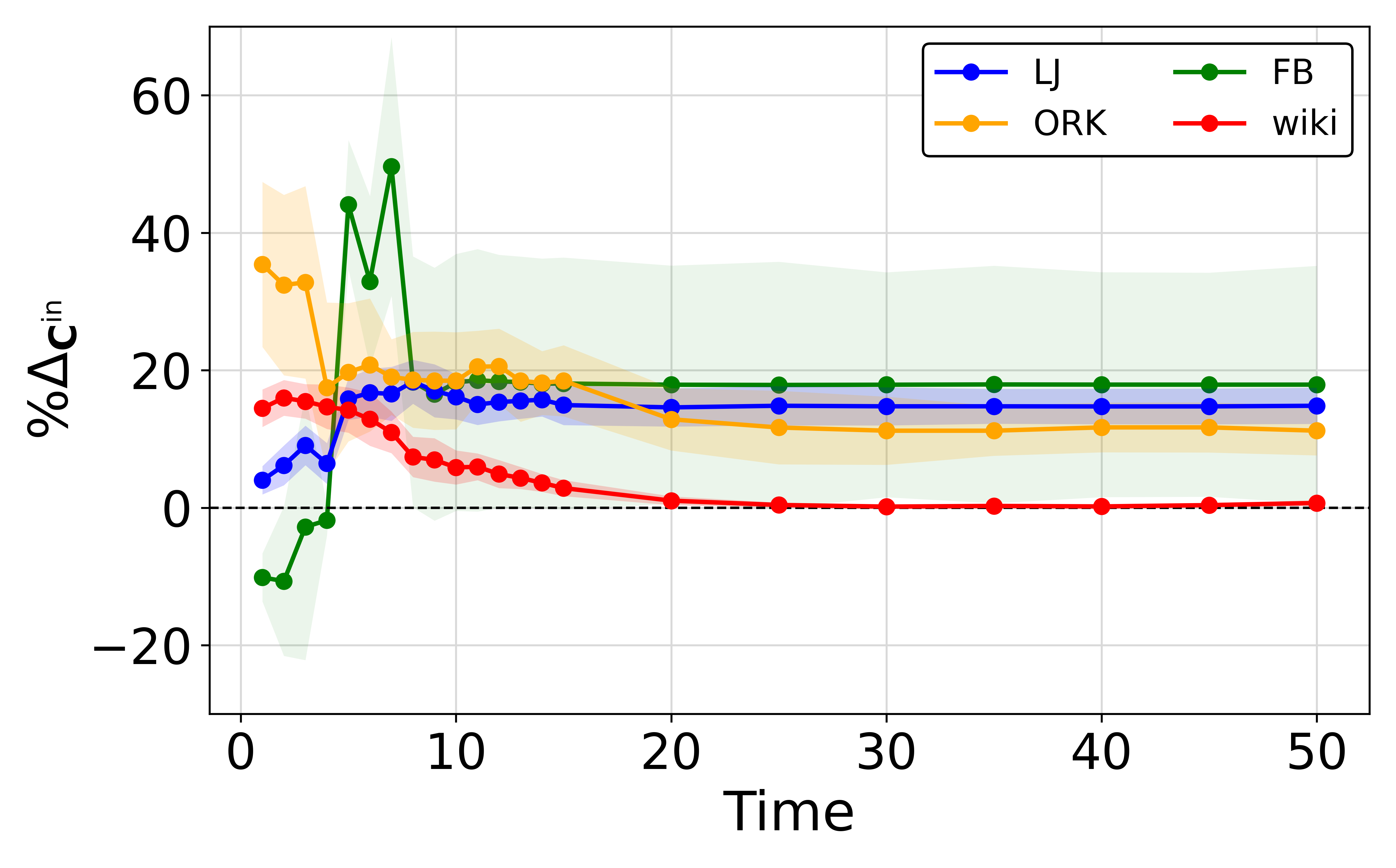}
    \caption{CC: In-Centrality}
    \label{fig:NEW-CC-IC}
\end{subfigure}

\vspace{0.5cm}

\begin{subfigure}[t]{0.48\linewidth}
    \centering
    \includegraphics[width=\linewidth]{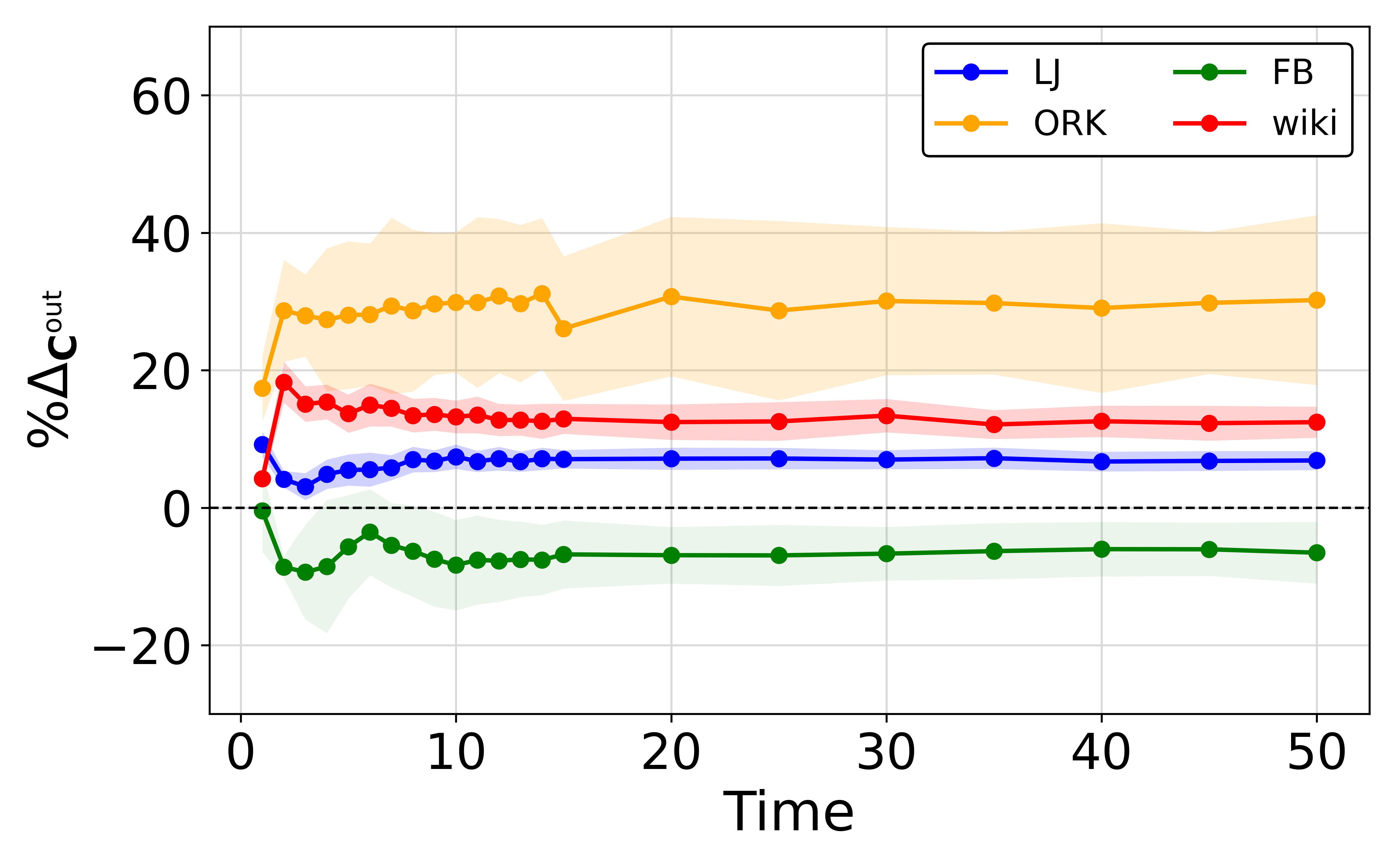}
    \caption{SC: In- and Out-Centrality}
    \label{fig:NEW-SC-OC}
\end{subfigure}

\caption{Temporal evolution of degree-controlled overlap effects across time (Eq. \ref{eq:betapercent}). Curves show dataset-level median relative change in centrality ($\%\Delta_{\textbf{C}^{\mathrm{in|out}}}$); shaded bands indicate $\pm 1$ standard error of the median. The top two panels report CC results for out-centrality (left) and in-centrality (right), while the bottom panel reports SC results for in- and out-centrality.}
\label{fig:degreecontrol-combined}
\end{figure}

To account for the effect of node degree, we use Equation \ref{eq:beta} to estimate the degree-controlled overlap effect. Figure \ref{fig:degreecontrol-combined} confirms that much of the variation in centrality is degree-driven. However, degree alone does not fully explain the observed centrality differences. The degree-controlled overlap effect remains non-zero across most networks, with the main exceptions occurring in Facebook networks under the SC model and during the early stages of CC spreading. For the rest of the datasets, the positive residual indicates that overlap structure contributes to centrality beyond simple differences in connectivity volume. More specifically, under the CC model, the degree-controlled effects remain positive for most networks and time points. In the diffused phase, the residual effect in Out-Centrality remains in the range of approximately 4\% to 33\%, with ORK dataset showing the largest residual effect. For In-Centrality, the effects are typically of the order of 20\%. However, in wiki networks the median effect declines toward zero as diffusion progresses, whereas in the other datasets the in-centrality difference remains close to 20\%, at a level comparable to that observed in Figure \ref{fig:stream_b}. For the SC model, the results are broadly similar to those of the CC model, consistent with the preceding analyses. After an initially variable phase, the diffused networks settle into a stable positive residual effect in out-centrality. 

Finally, to assess whether the degree-controlled overlap effects are systematically positive across networks, we apply Wilcoxon signed-rank tests to the network-level overlap coefficients at each diffusion time. The results support statistical significance for almost all diffusion times and datasets, with the main exceptions occurring in the Facebook networks; full p-value trajectories are reported in the Supplementary Material.

%
%
% TÄHÄN LOPPUU RESULTS
%
%
%
%
%
%
%
%4
%
% TÄSTÄ ALKAA DISCUSSION
%
%
%
%
\section*{Discussion}
In this section, we critically examine the methodological premises and analytical choices underlying our study. We aim to highlight potential biases and interpretative limits associated with commonly employed network modelling techniques and contrast them with our empirical approach. To do this we %We 
address four key aspects: the implications of rewiring methodologies, the phenomenon of diffusion saturation, the statistical rationale behind defining the analytical bulk of nodes, and the methodological considerations that guide the selection of circles. This is done to clarify how these decisions influence the interpretation of our results and the generalisability of conclusions regarding the overlapping structures and spreading dynamics in real-world networks.

\textit{Methodology.} Although rewiring methods are widely used in network analysis (e.g., in\cite{reid2011diffusion, shang2015epidemic}), rewiring methods could introduce biases in network analyses including distortions in the network topology, e.g., in terms of degree correlations and clustering coefficients \cite{Bertotti2020NetworkRewiring}. %. These biases include distortions in the network topology; for instance, degree correlations and clustering coefficients may change due to rewiring \cite{Bertotti2020NetworkRewiring}. 
Even if these structural characteristics are carefully maintained, the addition of edges may erase meso-scale structures, such as triadic closures that emerge naturally from homophily or attribute-based groupings \cite{karrer2008robustness}, and consequently essential structural features that allow reinforcement processes in complex contagions could be lost. Thus, observational methods based on empirical data are important in the analysis of real-world networks. Our analyses differ substantially from the rewiring experiments, because they reflect the natural 
emergence of overlaps due to social mechanisms, for example, by shared attributes or collective behaviour. A key concern is whether the observed advantage of overlapping nodes is merely a consequence of higher degree. Our degree-controlled analyses indicate that, while the  degree contributes to the spreading capacity, overlapping nodes retain a statistically significant advantage even when the degree is accounted for. This suggests that their role as connectors between multiple circles provides an additional structural benefit beyond the connectivity.

\textit{Saturation}. In \cite{jahani2022unequal}, the authors empirically demonstrated that information spreading frequently stalls when the links between real-world cross-group "broker" are insufficient or peripheral. Such scenarios result in uneven diffusion, even if average homophily measures suggest otherwise. In our study a similar indirect observation emerges, as low edge weights impede spreading, preserving the difference between OL and NOL nodes, even under saturated network conditions. Furthermore, without sufficient overlap involving central actors, entire communities can remain isolated, which is a phenomenon that could be invisible, e.g., in idealised rewiring scenarios. Empirical data highlight the important role of high-degree brokers, indicating that specific individual nodes, rather than the overall network topology, play a dominant role in the spreading. 

\textit{The Choice of the Bulk.} From statistical perspective, choosing quartiles (bottom 25\%, medium 50\%, top 25\%) for investigation is a poor choice, as the 
50\% bulk would represent too narrow a node range since the tail and the head of nodes' centrality distributions are heavily skewed. Typically, in real‑world networks, a majority of nodes are in the tail, whereas %the head-the most highly connected 
the most connected high degree nodes comprise only a small minority of the population \cite{newman2010networks}. Should we have chosen those limits, we would have observed the opposite results in the Lorenz curves, since the high end would have approximately exerted 60\% and medium  40\% of Betweenness Centrality (like the results obtained in \cite{alasadi2024clustering}). However, we can justify our limits by examining 
the cdf:s in Figure \ref{fig:ccdfs}: The largest gradient of the curves, i.e., the mass concentration of the nodes, is  
approximately within the shaded areas. Choosing the 50 or 60\% percentile cut-off would throw a portion of the true ‘medium’ nodes outside of the bulk and bias the analysis by excluding many of the nodes that actually concentrate most of the centrality mass, thus overstating the role of the extremes and misrepresenting the network’s spreading potential. We also recall that focusing solely on the ranking of centrality metrics does not necessarily predict the true spreading power. Our methods provide merely a probabilistic approach for the analysis and suggests %and %we can therefore predict \VK{this suggests} 
that because the bulk is responsible for most of the metric mass, they are more likely to contribute to spreading or initiate cascades.

\begin{figure}[ht]
  \centering
  \begin{subfigure}[t]{0.48\linewidth}
    \centering
    \includegraphics[width=\linewidth]{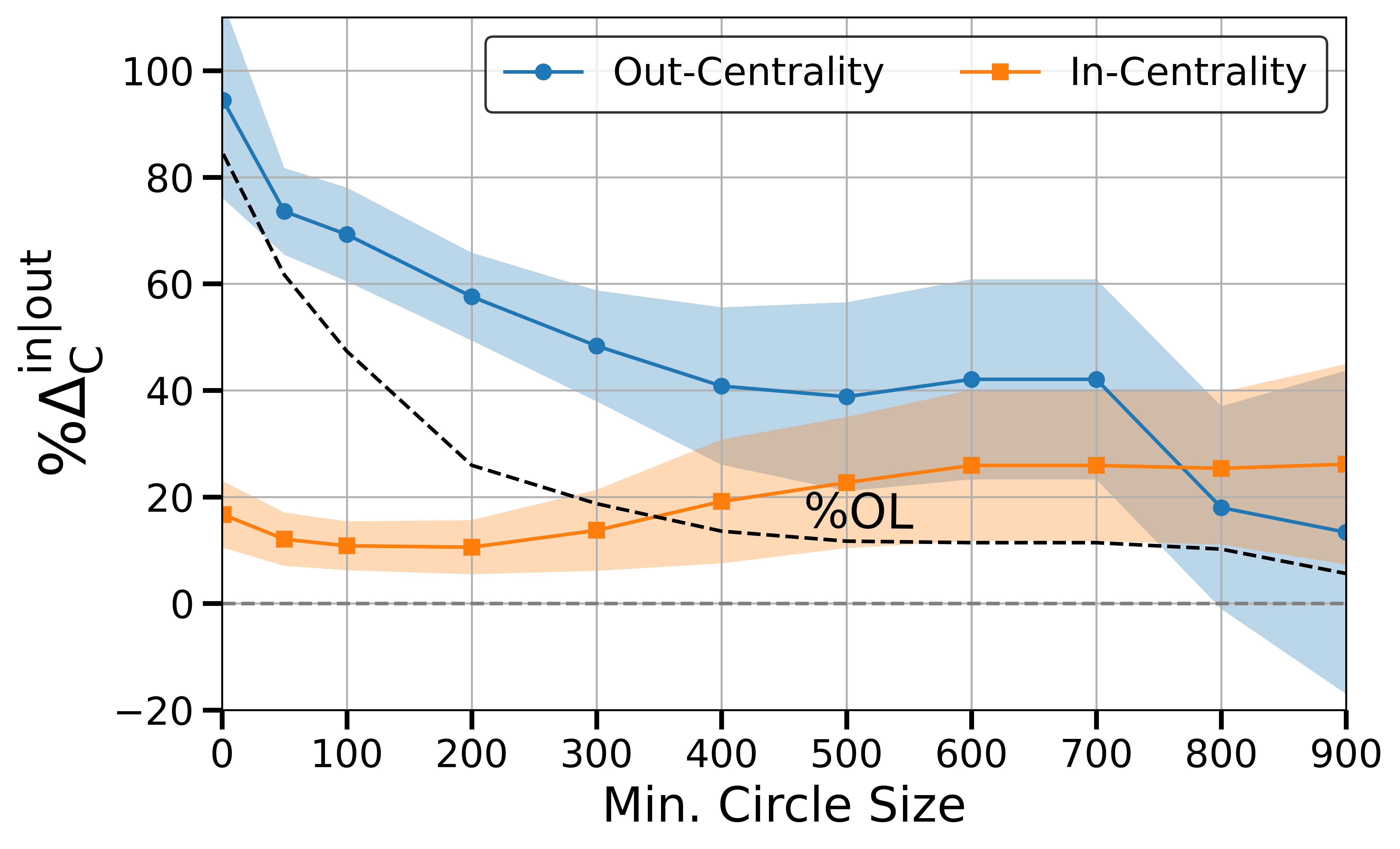}
    \caption{CC}
    \label{fig:cc_plot}
  \end{subfigure}
  \hfill
  \begin{subfigure}[t]{0.48\linewidth}
    \centering
    \includegraphics[width=\linewidth]{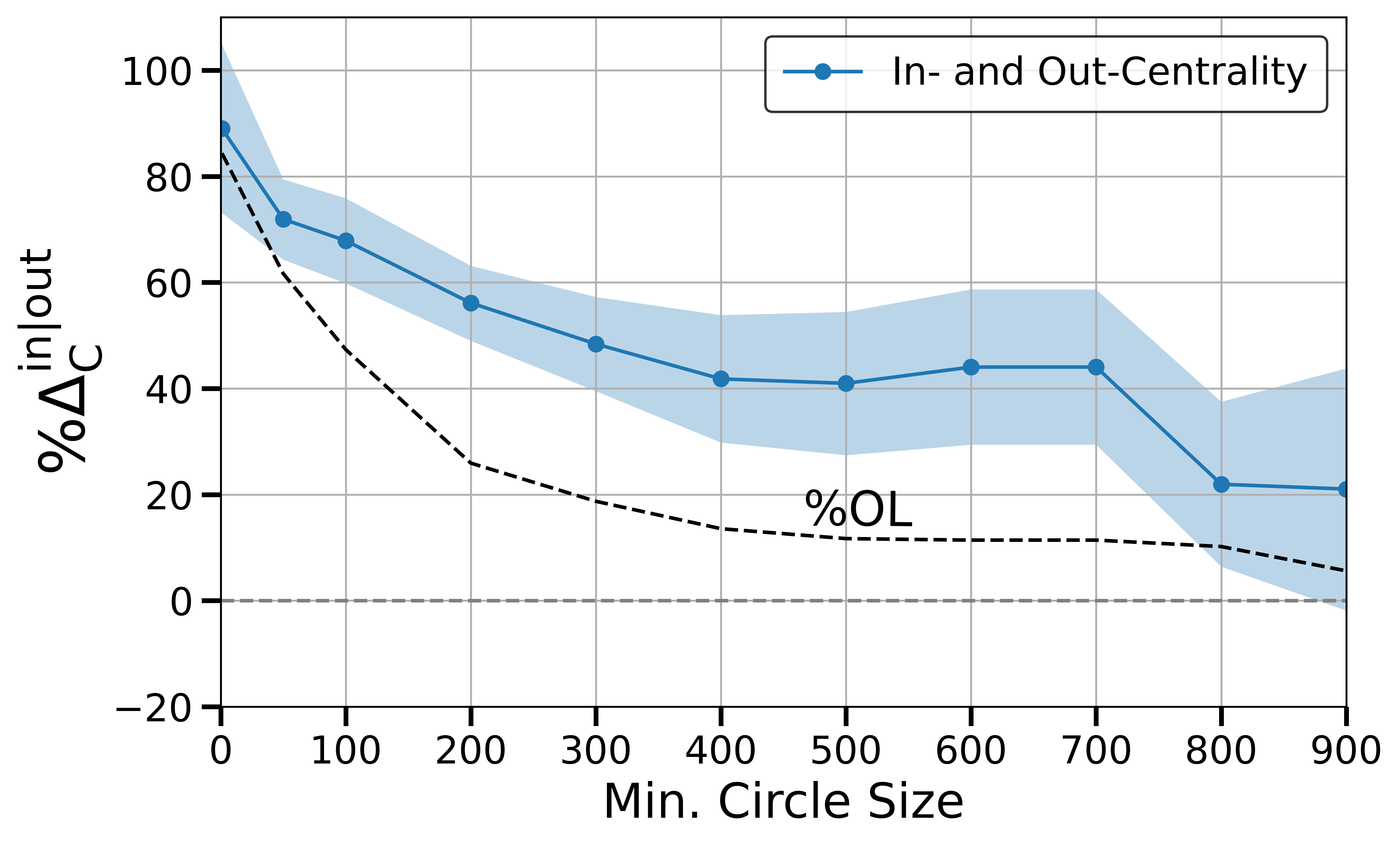}
    \caption{SC}
    \label{fig:sc_plot}
  \end{subfigure}
  \caption{ The OL and NOL relative difference in Out- and In-Centrality as a function of circle size for $T=100$ for the LJ dataset. The increase of circle's minimum size reduces the proportion of OL nodes. The most influential OL nodes reside in largest circles.}
  \label{fig:circles}
\end{figure}

\textit{The Choice of Circles.} Our analysis was guided by the rationale that allowing fewer nodes per circle would make the comparison of the OL and NOL nodes less reliable. Allowing smaller circles with fewer participants would capture a higher proportion of OL nodes and result in %much 
higher difference between node classes. Subsequently, only a few isolated and very peripherial nodes would be classified as NOL nodes and the whole comparison would become irrelevant. %We therefore restricted the minimum size of a circle to 10 nodes. 
Therefore, we %We therefore 
restricted the minimum size of a circle to 10 nodes to ensure meaningful structural comparisons. However, rather than interpreting larger circles as inherently more important, we examine how the node influence varies when restricting the analysis to progressively larger circles. Our results show that overlapping nodes retain their relative importance even under these constraints, thus suggesting that their influence is not solely driven by participation in small, dense substructures but persists across scales. Indeed, the importance of selecting circles becomes evident in the LJ dataset as shown in Figure \ref{fig:circles}. As smaller circles are progressively discarded and only larger circles are retained, more nodes are classified as NOL nodes. Although intuitively this might suggest a rapid convergence between the difference of metrics between OL and NOL nodes, our findings indicate the opposite. The importance and proportion of OL nodes decrease gradually, suggesting that while small circles and triadic structures capture important nodes initially, the key influencers %predominantly 
tend to reside within larger circles. The slight increase of the relative difference of In-Centrality also supports %backs up 
this claim. The most central nodes tend to have
susceptibility, even in diffuted networks. Our analysis consequently shows that removing the smallest circles preserves high influencers within the OL nodes. Thus, although circles reflect important topological properties of networks, identifying key nodes solely on the basis of overlaps may be insufficient. Therefore, we %We 
argue that additional community detection methodologies are necessary to identify effectively the true network influencers. % effectively. 
We predict that these influential nodes likely exist at the intersections of circles and community structures, and this would be one of the subjects of of our future work.

The analysis of circle size and the sustaining relative importance of OL nodes in largest circles raises an essential question: what constitutes an appropriate definition of a circle? For example, FB networks exhibit only a small percentage of OL nodes, suggesting that the current attributes used to form circles are inadequate to accurately capture all influential OL nodes and the structural properties of the networks. Conversely, datasets such as LJ and ORK show that a substantial proportion of nodes are captured within circles if we take into account even the smallest circles of three nodes. In those datasets, group memberships provide a natural basis for defining overlapping structures; In contrast, in datasets such as FB, attribute-based circles (e.g., friend lists) may not correspond closely to structurally cohesive subgraphs. Our results reflect this variation. In particular, the relative differences between OL and NOL nodes are less pronounced in datasets with a low proportion of overlapping nodes or less structurally informative attributes. This suggests that the observed effects depend on how well the attribute-based circles align with underlying network structure. See Appendix A for further analysis regarding the choice of circles.
%
%
% end of discussion
%
%
%
%
%
%
%
%
%
%
%
% begin of conclusions
%

\section*{Conclusions}
We have presented a comparative analysis of the importance of nodes in overlapping circles using metrics derived from the probabilistic Influence Spreading Model. By comparing the spreading properties of overlapping and non-overlapping nodes through both distributional and temporal analyses, we show that, within the studied datasets and modelling assumptions, overlapping nodes exhibit higher relative influence across a range of network settings. This advantage is visible in their susceptibility, spreading power, and mediating role, as captured by the selected centrality metrics. In addition, our degree-controlled analysis shows that although degree explains a substantial part of the observed variation, the advantage of overlapping nodes often persists after accounting for degree, indicating that overlap is associated with an additional structural role beyond connectivity alone. We further find that not only the top influencers but also the central bulk of nodes contribute substantially to the spreading process for the chosen centrality metrics. Our analyses also reveal that overlapping nodes exhibit shifted distributions and, on average, attain higher values across the selected metrics than their non-overlapping counterparts.

Our findings show that although the studied networks differ structurally, their overall behaviour during influence spreading remains similar for both complex-contagion and simple-contagion models. More precisely, the relative In-Centrality shows a consistent decline, indicating higher initial susceptibility for overlapping nodes, which eventually stabilises. Conversely, the Out-Centrality highlights that overlapping nodes retain substantial spreading power even during the diffused phases. Furthermore, subtle but meaningful features, such as accumulation periods, emerged in our metric analysis. We discovered an initial rise in the relative importance of overlapping nodes. Subsequently, the Betweenness Centrality revealed a delayed temporal evolution, showing that the overlapping nodes can retain their mediating role for a longer period in the spreading process.

In addition, we investigated how the definition and selection criteria for circle attributes shape the importance of a node.  
Although restricting the size of the circle slightly reduces the relative importance of overlapping nodes,  
the reduction occurs gradually, which implies that the largest circles predominantly host super-influenting nodes. This observation emphasises that the overlapping nodes are %a key part of 
closely linked to important topological properties, such as triadic closures and cliques, which are important natural structures of real complex networks. Our analysis, put together, %strongly supports 
supports the potential %the 
strategic utilisation of overlapping nodes. For example, in cybersecurity networks, overlapping nodes could help detect and mitigate vulnerabilities and serve as points for proactive security interventions. Future work could extend our analysis by using topology-based community detection alongside attribute-based circles, enabling a more comprehensive identification of structurally important nodes. Our goal is to identify and isolate nodes located at the intersections of both overlapping communities and overlapping circles, and distinguish the most essential nodes of the networks.

%\noindent LaTeX formats citations and references automatically using the bibliography records in your .bib file, which you can edit via the project menu. Use the cite command for an inline citation, e.g.  \cite{Hao:gidmaps:2014}.

%For data citations of datasets uploaded to e.g. \emph{figshare}, please use the \verb|howpublished| option in the bib entry to specify the platform and the link, as in the \verb|Hao:gidmaps:2014| example in the sample bibliography file.

\newpage
\section*{Appendix}
\subsection*{Appendix A. Synthetic Circles}
Attributes do not always provide the possibility of directly identifying overlapping circles from data. On platforms like Facebook, overlapping occurs naturally when users (nodes) belong to multiple groups. In Wikipedia's top 100 categories, overlapping arises because pages often belong to multiple categories and thus fall into multiple circles. When establishing node attributes, one approach is to define overlapping circles through intersections of attribute values. For example, combining two distinct attributes--such as "height" and "age"--allows the creation of circles representing users who share these properties. A node would then be considered overlapping if it belonged simultaneously to both "age" and "height" groups. Conversely, if a user chooses not to disclose age, and only another attribute is available, the node would be classified as non-overlapping because it belongs to a single group only. We employed this logic in our analysis. 

\begin{figure}[H]
    \centering
    \includegraphics[width=0.5\linewidth]{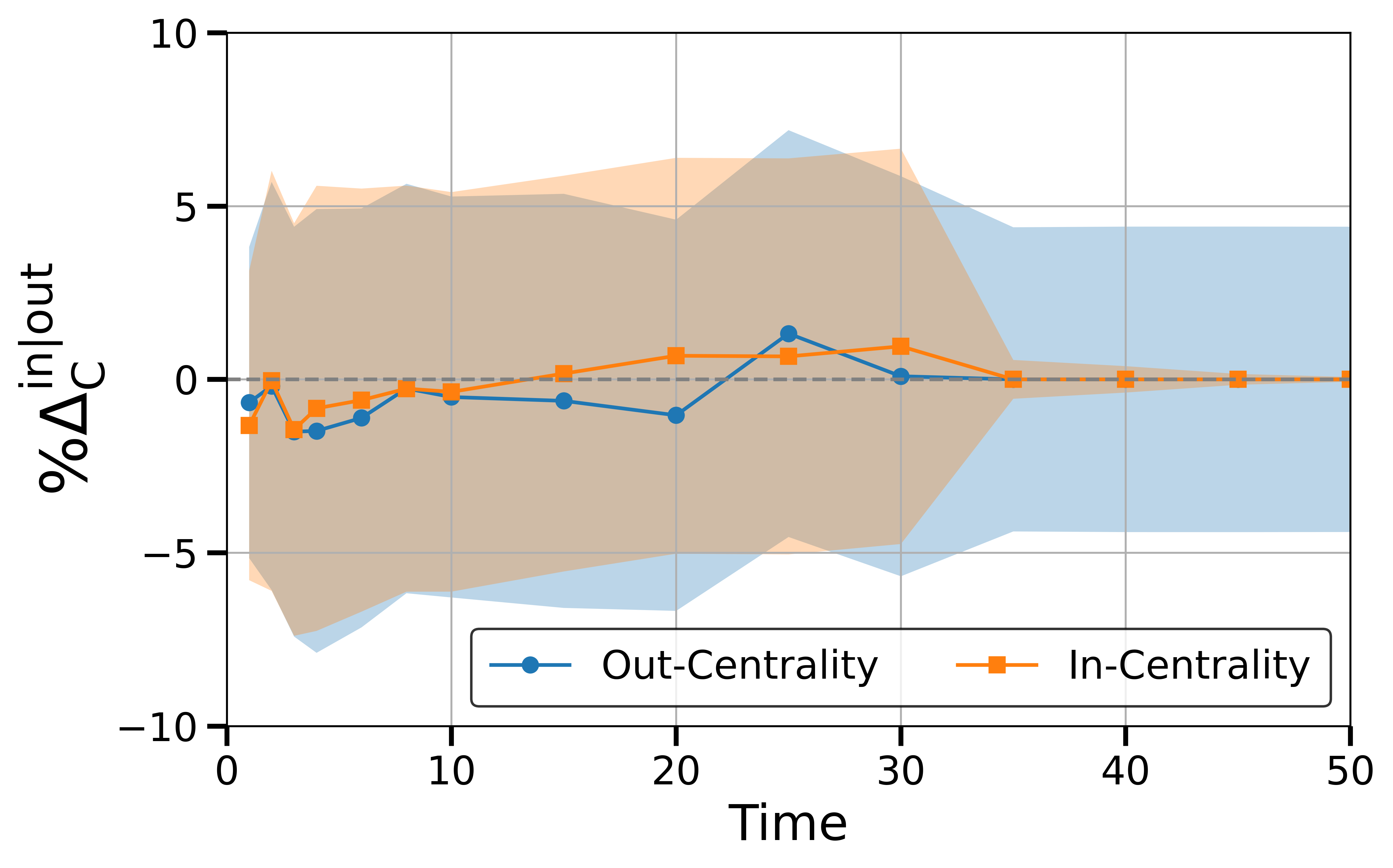}
    \caption{Complex Contagion with Pokec data. No relative difference in centralities when deriving synthetic circles.}
    \label{fig:Pokec}
\end{figure}

Specifically, we examined the Pokec dataset\cite{takac2012data}, a social network analogous to other datasets used in our experiments (see Table \ref{tab:Pokec} for details). Unlike the LiveJournal dataset, where overlapping circles are formed from users' memberships in multiple user-defined groups, the Pokec dataset does not contain a single attribute that creates similar overlaps. For instance, "Region ID" attribute represents the user's home region, and Pokec does not permit the selection of multiple regions, which would naturally allow overlapping circle formation. Therefore, to introduce overlaps, we combined the attributes "Region ID" and "Age," defining overlapping properties for each unique combination of region ID and age values. Users sharing either the same region ID or the same age value belong to the corresponding circles. Users with missing information in either "Region ID" or "Age" were classified as non-overlapping. In real-world scenarios, users sharing only an age but residing in different regions are unlikely to have meaningful social connections. This observation aligns with our empirical analysis: the relative differences between In- and Out-Centrality measures among overlapping and non-overlapping nodes were negligible. Figure \ref{fig:Pokec} illustrates these findings. This observation reinforces our observations that genuinely overlapping nodes have topological importance and are likely related to triad or clique formations within the network. Furthermore, attributes associated with overlapping nodes are typically non-random and demonstrate a strong correlation with other attributes. This phenomenon is known as homophily.

\begin{table}[H]
    \centering
    \begin{tabular}{c c c c c}
        \hline
        Networks & Nodes & Clustering & Average Degree & Overlapping \\ \hline
        21 & $888_{715}^{5416}$ & $0.20_{0.09}^{0.71}$ & $6.9_{3.20}^{21.7}$ & $66.3_{46.2}^{77.6}$ \\ 
       \hline
    \end{tabular}
    \caption{Average, minimum and maximum properties of Pokec ego-networks derived from dataset.}
    \label{tab:Pokec}
\end{table}

\subsection*{Appendix B. Choosing the Edge Weights}
\begin{figure}[ht]
  \centering

  % First row
  \begin{subfigure}[t]{0.48\linewidth}
    \centering
    \includegraphics[width=\linewidth]{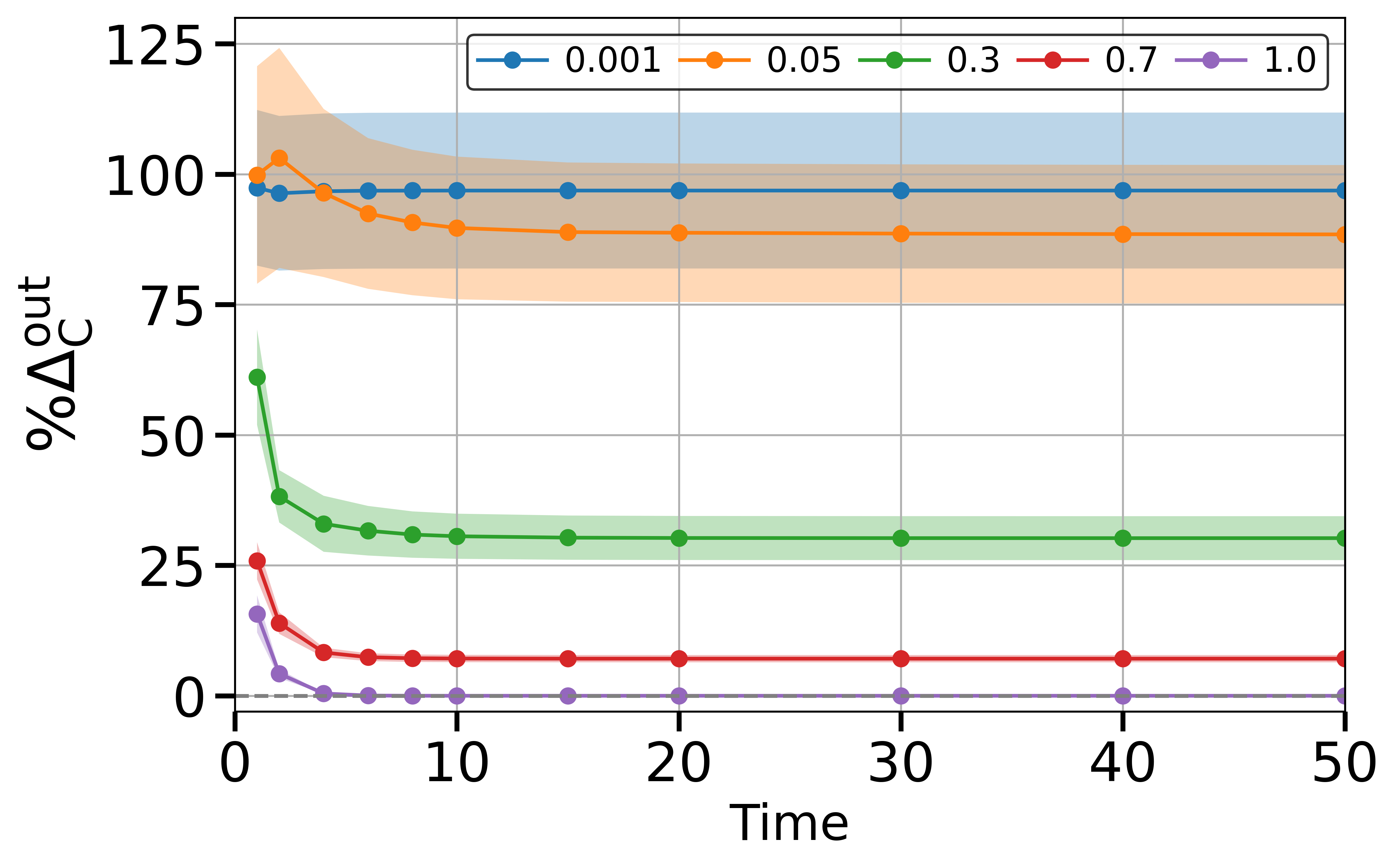}
    \caption{CC: Out-Centrality}
    \label{fig:cc_0.3}
  \end{subfigure}
  \hfill
  \begin{subfigure}[t]{0.48\linewidth}
    \centering
    \includegraphics[width=\linewidth]{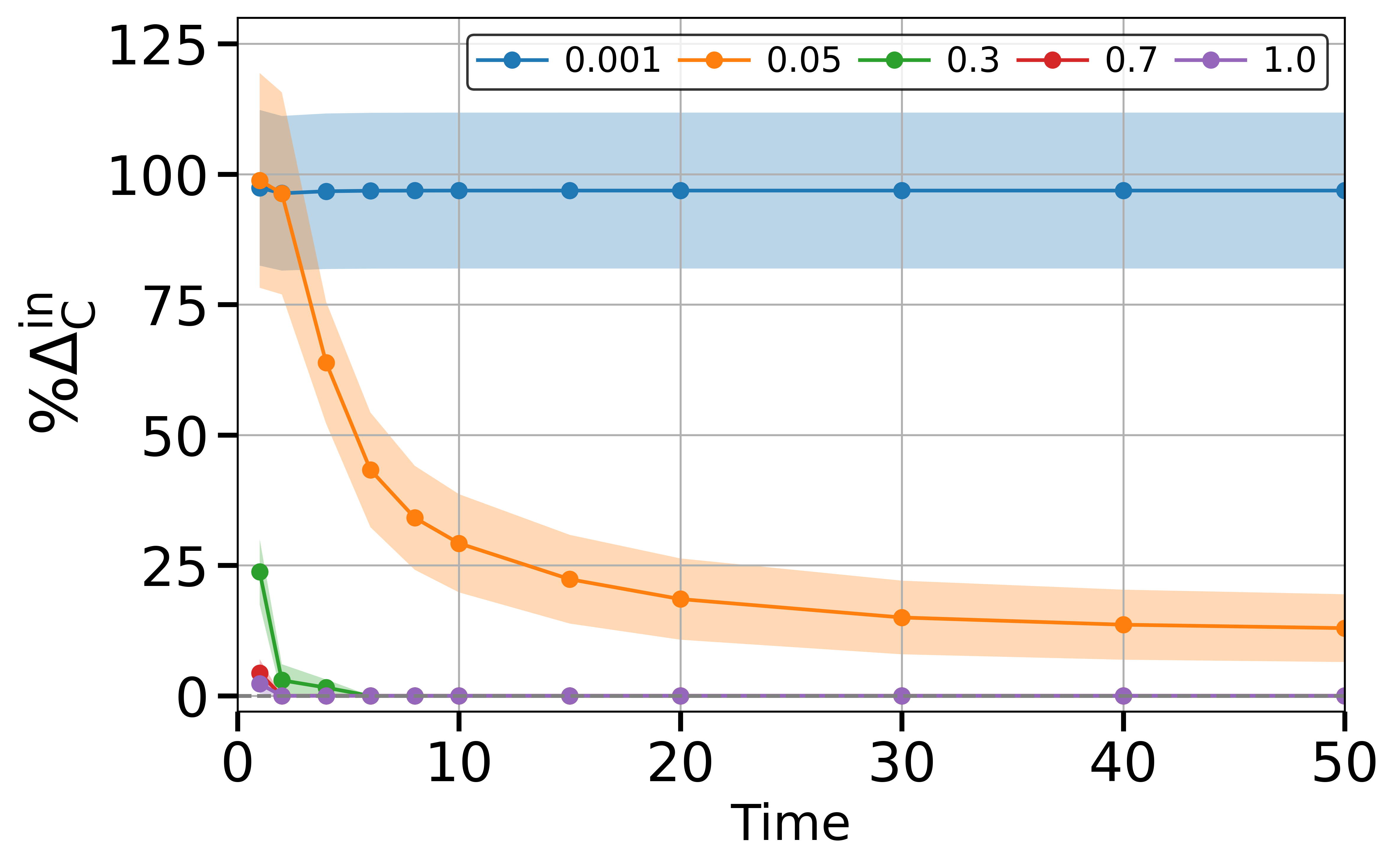}
    \caption{CC: In-Centrality}
    \label{fig:cc_0.7}
  \end{subfigure}

  \vspace{0.5cm}

  % Second row
  \begin{subfigure}[t]{0.48\linewidth}
    \centering
    \includegraphics[width=\linewidth]{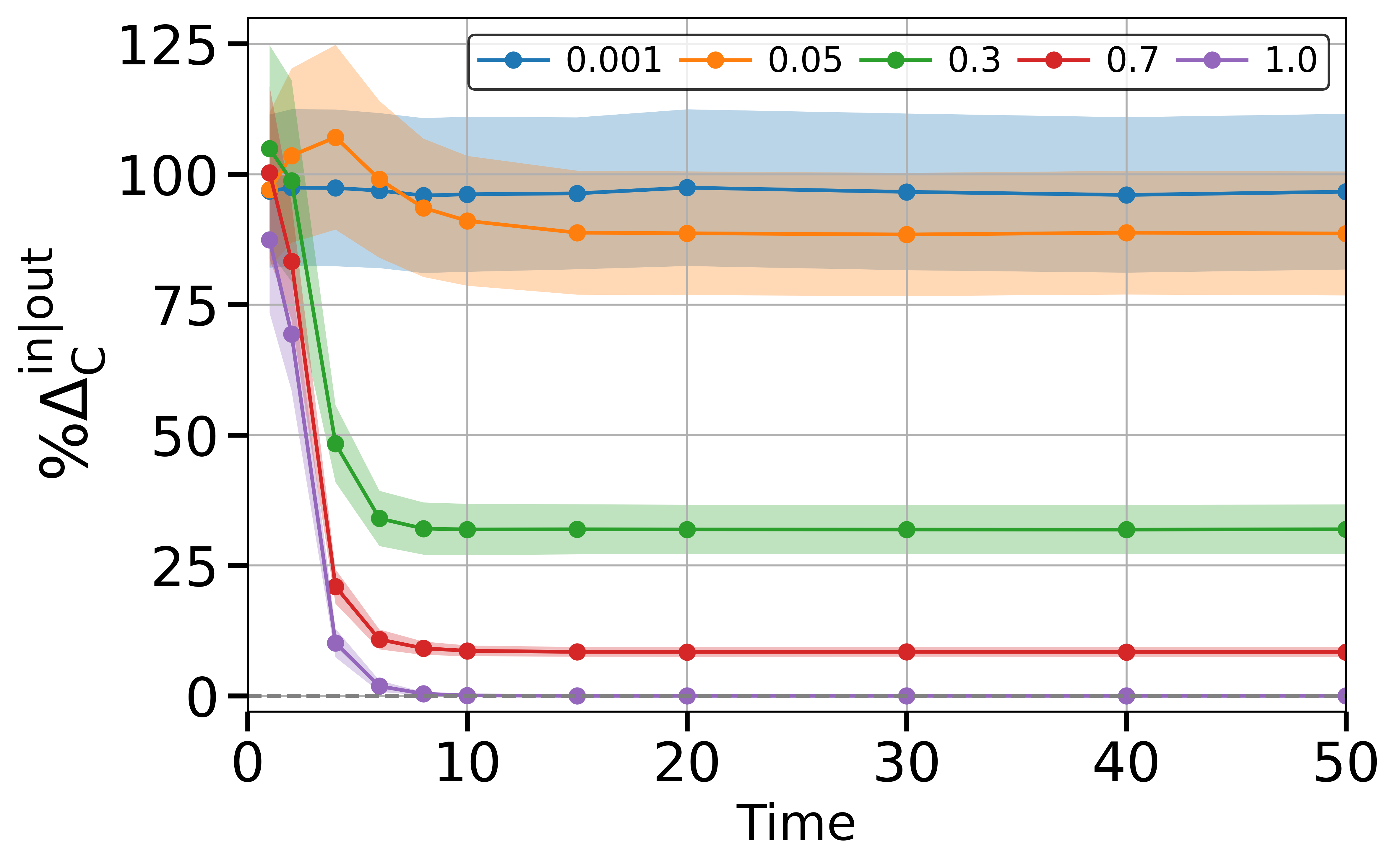}
    \caption{SC: In- and Out-Centrality}
    \label{fig:sc_0.3}
  \end{subfigure}
  \hfill

  \caption{The OL and NOL relative difference in both complex (top) and simple contagion (bottom) as a function of time for ORK networks with various weights. In- and Out-Centralities are, again, equivalent for SC, which is a property for undirected networks.}
  \label{fig:orc_combined}
\end{figure}

It is well-known edge weights influence the most to information passing in networks \cite{bellingeri2023considering}. Too small edge weights hold the spreading contained, while too large weights rapidly saturate the network. Therefore, we re-ran our analysis with different uniform edge weights to ensure the differences between overlapping and non-overlapping nodes exist; that they are not just a bias due to low edge weights. We performed the test for ORK datasets' ego-networks with uniform weights $0.001, 0.05, 0.3, 0.7$ and $1.0$, holding the rest of the parameters the same as in previous experiments. The results are shown in Figure~\ref{fig:orc_combined}. The difference between overlappers and non-overlappers remains, although the difference evens up with higher weights. This is because at high weights the transmission probability per contact approaches 1, so cascades propagate across every edge, and the positional advantage of overlappers diminishes. The weights ($0.05$) used in the study, however, yield the slow enough spreading for capturing the gradual saturation, and even some accumulation.

The SC-model is less sensitive to weight alteration in the beginning of contagion, while with the CC-model the difference diminishes as the edge weights close towards 1. Both models show only minor differences in the saturated phase. Furthermore, spreading stabilises faster with higher weights, as the information is allowed to pass more likely through the edges. On the contrary, with very low edge weights, the information cannot pass through the network, and the relative difference between overlapping and non-overlapping nodes remains high, even though the spreading remains weak. For the purposes of this study, the edge weight $0.05$ is suitable for examining the smooth decline of In-Centrality, which does not occur with lower weights, for example, with $0.001$. Furthermore, the low-weight setting allows for examining the spreading in more detail in the beginning of the simulation. A more accurate resolution would be obtained either by choosing larger networks or by increasing the cadence of observations.

\bibliographystyle{unsrt}
\bibliography{sample}

\end{document}